\documentclass[lettersize,journal]{IEEEtran}
\usepackage{amsmath,amsfonts}
\usepackage{algorithmic}
\usepackage{algorithm}
\usepackage{array}
\usepackage{textcomp}
\usepackage{stfloats}
\usepackage{url}
\usepackage{verbatim}
\usepackage{graphicx}
\usepackage{cite}
\usepackage{hyperref}
\hypersetup{hidelinks}
\usepackage{multirow}
\usepackage{utfsym}
\usepackage{colortbl}
\usepackage{makecell}
\usepackage{booktabs}

\usepackage{pifont}
\newcommand{\cmark}{\ding{51}}%
\newcommand{\xmark}{\ding{55}}%

\usepackage{CJKutf8}

\newcommand{\gain}[1]{\textcolor{green}{#1}}
\newcommand{\drop}[1]{\textcolor{red}{#1}}

\newcommand{\re}[1]{\textcolor{black}{#1}}

\newcommand{\eg}{\emph{e.g.}~}

\newcommand{\etal}{\emph{et al.}~}
\newcommand{\ie}{\emph{i.e.}~}

\begin{document}
\begin{CJK*}{UTF8}{gbsn}


\title{ASR-enhanced Multimodal Representation Learning for Cross-Domain Product Retrieval}

\author{Ruixiang Zhao, Jian Jia, Yan Li, Xuehan Bai, Quan Chen, Han Li, Peng Jiang, Xirong Li, ~\IEEEmembership{Member,~IEEE,}
\thanks{Manuscript received August 6, 2024; revised December 27, 2024 and June 20, 2025; accepted June 22, 2025. This work was funded by NSFC (No. 62172420) and Kuaishou. The associate editor coordinating the review of this article and approving it for publication was Prof. Min Chen.  \emph{(Corresponding author: Xirong Li.)}}
\thanks{Ruixiang Zhao and Xirong Li are with Renmin University of China, Beijing 100872, China  (e-mail: \{ruixiangzhao, xirong\}@ruc.edu.cn).}
\thanks{Jian Jia, Yan Li, Xuehan Bai, Quan Chen, Han Li and Peng Jiang are with the Kuaishou Technology, Beijing 100085, China (e-mail: \{jiajian,  liyan26, baixuehan03, chenquan06,  lihan08, jiangpeng\}@kuaishou.com).}}


\markboth{IEEE TRANSACTIONS ON MULTIMEDIA,~Vol.~?, No.~?, ?~2025}%
{Shell \MakeLowercase{\textit{et al.}}: A Sample Article Using IEEEtran.cls for IEEE Journals}

\IEEEpubid{0000--0000/00\$00.00~\copyright~2025 IEEE}

\maketitle

\begin{abstract}
  E-commerce is increasingly \emph{multimedia}-enriched, with products exhibited in a broad-domain manner as images, short videos, or live stream promotions. A unified and vectorized cross-domain production representation is essential. Due to large intra-product variance and high inter-product similarity in the broad-domain scenario, a visual-only representation is inadequate. While Automatic Speech Recognition (ASR) text derived from the short or live-stream videos is readily accessible, how to de-noise the excessively noisy text for multimodal representation learning is mostly untouched. We propose \underline{A}SR-enhanced \underline{M}ultimodal \underline{P}roduct R\underline{e}p\underline{r}esentation L\underline{e}arning (\texttt{AMPere}). In order to extract product-specific information from the raw ASR text, \texttt{AMPere} uses an easy-to-implement LLM-based ASR text summarizer. The LLM-summarized text, together with visual data, is then fed into a multi-branch network to generate compact multimodal embeddings. Extensive experiments on a large-scale tri-domain dataset verify the effectiveness of \texttt{AMPere} in obtaining a unified multimodal product representation that clearly improves cross-domain product retrieval. 
\end{abstract}


\section{Introduction} \label{sec:intro}
\begin{figure}[!t]
    \centering
    \begin{minipage}[c]{\linewidth}
        \centering        \includegraphics[width=\textwidth]{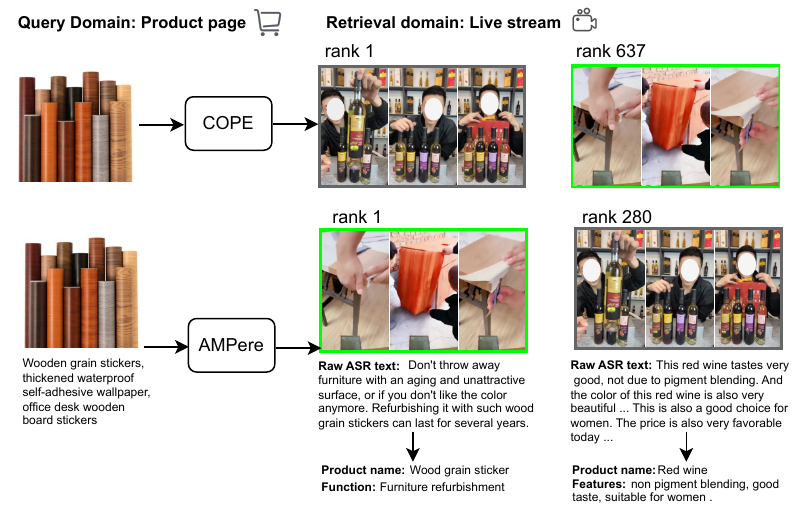}
        \centerline{(a) Product: \textit{Wood grain sticker}} \label{fig:wood}
    \end{minipage}

    \vspace{2mm}
    \begin{minipage}[c]{\linewidth}
        \centering
        \includegraphics[width=\textwidth]{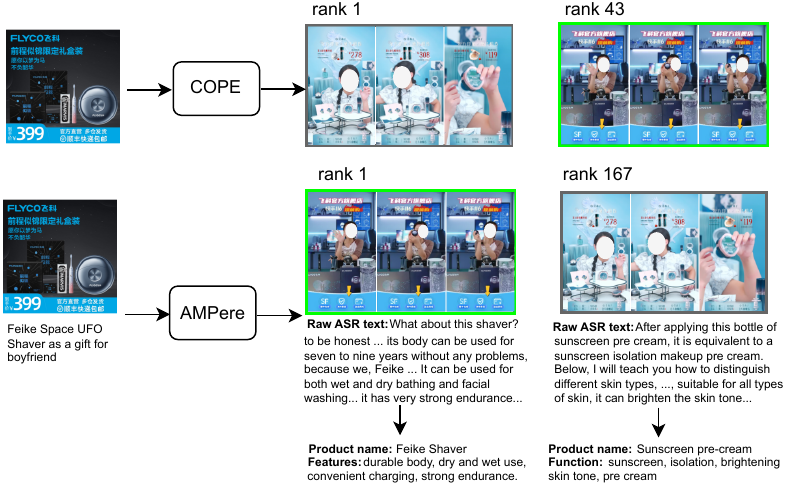}
        \centerline{(b) Product: \textit{Electric shaver}}
    \end{minipage}
    
    \caption{\textbf{Exemplars of cross-domain product retrieval (CdPR)}. Here a query of a specific product, \ie (a) wood grain sticker and (b) electric shaver, from the product-page domain is used to search for live-stream videos selling the product. CdPR is challenging due to large intra-product variance and high inter-product similarity. Compared to COPE \cite{rope}, the latest \emph{visual} based solution, our proposed \texttt{AMPere} method, with its novel use of Automatic Speech Recognition (ASR) text for \emph{multimodal} product representation learning (MmPRL), produces much-improved retrieval results. Note that our experimental data is in Chinese. We translate the text to English for non-Chinese audience.}
    \label{fig:showcase}
\end{figure}

\IEEEPARstart{T}{he} content of E-commerce is becoming increasingly \emph{multimedia}-enriched. Besides traditional product pages, E-commerce platforms are incorporating more dynamic and interactive media formats, such as short videos and live streams, to present their products more effectively and consequently enhance end-user shopping experience. A specific product can now be exhibited in a broad-domain manner through an image \cite{gu2018multi}, a short descriptive video \cite{real20m}, or a live stream promotion \cite{rope}, see Fig. \ref{fig:showcase}. As such, a unified and vectorized cross-domain representation of the product is essential for downstream tasks including cross-domain product retrieval (CdPR) \cite{rope}, user profiling, personalized recommendation, \re{and online video advertising \cite{videoecommerce, videoecommerce++}, to name just a few}. This paper is targeted at \emph{\textbf{m}ulti\textbf{m}odal} \textbf{p}roduct \textbf{r}epresentation \textbf{l}earning (MmPRL) for CdPR.

Current efforts in product representation learning are primarily focused on the product-page domain \cite{pang2022heterogeneous,fashionbert,kaleidobert,eiclip,adscvlr,fashionvil,commercemm,date}, where a product sample appears as an image-title pair. EI-CLIP \cite{eiclip}, for instance, utilizes a CLIP \cite{clip} based network to encode the title and the image into two cross-modal embeddings. 
\IEEEpubidadjcol
AdsCVLR \cite{adscvlr} produces a multimodal embedding for the image-text pair with a transformer architecture. Gu \etal \cite{gu2018multi} propose a multimodal and multi-source embedding learning framework for fashion retrieval. All the above works \cite{eiclip,adscvlr,gu2018multi} consider the image domain only. In contrast, larger intra-product variance exists when searching for products across image and video domains. As exemplified in Fig. \ref{fig:wood}, for products such as stickers, the product image differs substantially from the visual appearance of stickers in use. Product representation learning in the multi-domain setting is thus more challenging. 

To attack the challenge, Bai \etal develop a tri-domain benchmark termed ROPE\footnote{\href{https://github.com/adxcreative/COPE}{https://github.com/adxcreative/COPE}}, containing millions of multimodal samples from three domains, \ie product-page, short-video and live-stream \cite{rope}. Their proposed solution for CdPR is fully reliant on the visual content, which alone is insufficient to handle large intra-product variance and high inter-product similarity, see Fig. \ref{fig:showcase}. Chen \etal propose a query-driven approach \cite{real20m}, assuming that a product image or a short video is associated with a textual query based on click-through records. However, such a text modality is often unavailable in particular for novel products. In a contemporary study, Yang \etal propose to exploit Automatic Speech Recognition (ASR) text transcribed from the audio channel of short / live-stream videos  \cite{rice}. Despite its wide availability, the ASR text is typically overwhelmed by abundant uninformative words from casual chatting of network
anchors or live streamers, see Table \ref{tab:llm_prompt}. Indeed, the raw ASR text is previously deemed to be harmful for product representation \cite{rope}. 

\begin{table}[tbp]
\caption{\textbf{Our prompt to let an LLM} (Baichuan2-13B-Chat-4bits) \textbf{perform ASR text summarization}. We experiment with Chinese data, so the prompt used is written in Chinese. Its English counterpart is shown for non-Chinese readers. \label{tab:llm_prompt}} 

\renewcommand{\arraystretch}{1}
\resizebox{\linewidth}{!}{
\begin{tabular}{p{9.5cm}}
\toprule

\rowcolor{gray!20} You are a text cleaning expert who needs to clean and streamline the words spoken by an anchor during live-streaming sales. Need to remove redundant information and persuasive words, and retain only objective information related to the product. Please carefully observe the examples below and output in the specified format.\\ 
\\
Example:\\
\textbf{Input}: \emph{好了199的价格做起来 我们家的1号链接感谢大家的支持啊 ... 把你们想看链接报出来我们公屏上主播拿给大家讲一下也是可以的好不好 小飞碟这款递须刀呢拿去送给男朋友送老公非常的合适啊它是个便携款而且呢是一个精致小巧的款式啊精致小巧的款式 随身携带呢是可以做到的啊非常非常非常便捷款标 而且呢刀头材质比较做的非常好所以说体验感是非常棒的啊 弧面双环的网54度的0.35毫米纹刀片又薄又润刀片你们去剃须手是不费力的啊 只需要十五秒钟到二十秒钟时间快速剃须而且胡子长的胡子硬的胡子粗的胡子重的都能给大家剃干净 不留小黑点非常光溜啊包括这款妆是个呃刀头是我们现在市面上新刀头 越磨越丰利越磨越丰利用的时间越长是越经济越实惠的五到八年不用换到片五到八年不用大家去换到头随便用啊 非常的经济续航时间多长 充电一个小时可以达到60天的超长需要时间啊60天超需要时间下面这个充电口子呢是一个通用的充电口子啊通用的充电口子}\\

\underline{Translation purely for readability}: \emph{Alright, let's start with the price of 199. Thank you for your support on our link number 1. This space UFO shaped shaver is very suitable for giving to boyfriends and husbands. It's a portable and delicate style that can be carried with you. It's very, very convenient. The material of the blade head is very good, so the experience is very good. The curved double-ring net with 54 degrees and 0.35mm stripe blade is thin and moist. It only takes fifteen to twenty seconds to shave quickly, and any beard that is long, hard, thick, or heavy can be shaved clean without leaving small black spots. It is very smooth, including this makeup. The blade is a new blade on the market, and the more it is sharpened, the more effective it is. The longer it is used, the more economical and affordable it is. It does not need to be replaced for five to eight years. You do not need to change the blade for five to eight years. You can use it freely. It is very economical. The battery life is longer. Charging for one hour can reach 60 days, which takes more time. The following charging port is a universal charging port. Universal charging port.}\\ [3pt]
\textbf{Output}:\\
\texttt{Product name}: \emph{Space UFO shaped shaver}\\
\texttt{Features}: \emph{Price 199 yuan, portable style, exquisite and compact; The blade material is of high quality and provides a good user experience; Curved double ring mesh, shaving is fast and clean; The more you apply, the more effective it becomes, and the longer you use it, the more economical it becomes; Easy to charge, long battery life, universal charging port.}\\
\ldots
\\
\rowcolor{gray!20}
\textbf{Input}: \$\{raw ASR text\}\\
\textbf{Output}: \\
\texttt{Product name}: \$\{name of the product\}\\
\texttt{Features}: \$\{major features of the product\}\\
\bottomrule
\end{tabular}
}
\end{table}

\begin{figure*}[htbp]
    \centering
    \includegraphics[width=0.87\linewidth]{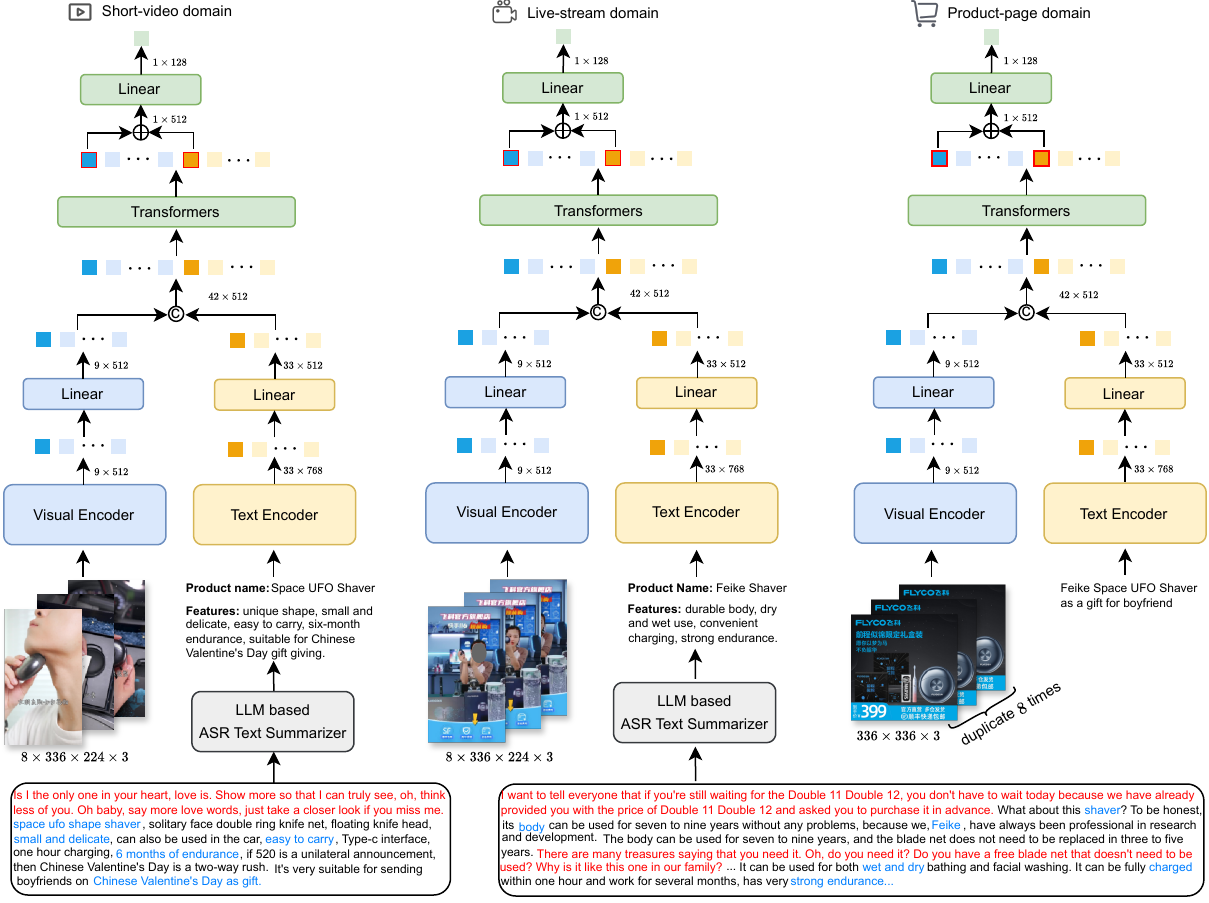}
    \caption{\textbf{Proposed \texttt{AMPere} method for cross-domain multimodal product representation learning}. Our network consists of three branches, each responding to a specific domain, \ie \emph{short-video}, \emph{live-stream} and \emph{product-page}. For each trainable layer in the network, its parameters are shared across the three branches. An LLM-based ASR text summarizer is deployed to extract \textcolor{blue}{product-specific information}  from the raw text overwhelmed by \textcolor{red}{abundant uninformative words}  transcribed from casual chatting of network anchors or live streamers. Except for the summarizer, the network is trained end-to-end.}
    \label{fig:framework}
\end{figure*}

Although ASR text has been used in varied video tasks such as text-video retrieval \cite{liu2019CE}, video representation learning \cite{howto100m,aaai25-VDSFX}, video abstractive summarization \cite{sffg},video question answering \cite{zellers2021merlot} \re{and video hyperlinking \cite{hyperlinking, cheng2017selection}}, how to de-noise the text for multimodal representation learning is almost untouched. Given the excessive noise in ASR text derived from E-commence videos, we empirically find that an advanced information extraction tool as UIE \cite{uie} fails to extract anything for over 30\% of the ROPE data. Inspired by the great success of large language models (LLMs), we propose to de-noise the ASR text with an LLM-based text summarizer, see Tab. \ref{tab:llm_prompt}. MmPRL with the enhanced ASR text leads to much improved CdPR results.
In short, our major contributions are as follows:
\begin{itemize}
    \item We propose \underline{A}SR-enhanced \underline{M}ultimodal \underline{P}roduct R\underline{e}p\underline{r}esentation L\underline{e}arning (\texttt{AMPere}), see Fig. \ref{fig:framework}, showing for the first time that incorporating ASR text can largely benefit product retrieval in the E-commerce domain.
    \item We propose an easy-to-implement LLM-based ASR text summarizer, which can effectively extract product-specific information from the verbose and extremely noisy text. LLM-based text summarization is by itself not new. Our key novelty lies in the novel use of this technique to successfully unlock the value of the ASR text, previously recognized unusable, for MmPRL.
    \item Extensive experiments on the large-scale ROPE dataset verify the viability of the proposed method for obtaining a unified multimodal product representation for CdPR.
\end{itemize}

\section{Related Work} \label{sec:rel}
This work is targeted at learning ASR-enhanced multimodal production representation for CdPR. Hence, we first discuss recent progress in CdPR, followed by a brief survey concerning the use of ASR text in a broader context.

\subsection{Cross-domain Product Retrieval}

\re{Despite good attempts on cross-domain image-to-image retrieval in a non-ecommerce context \cite{pros} and} single-domain (mostly product-page) product retrieval \cite{fashionbert,kaleidobert,eiclip,fashionvil,adscvlr,commercemm,date,liu2023mep,m5product}, there are relatively few studies on cross-domain product retrieval, see Table \ref{tab:re_cdpr}. 
\re{In an earlier work \cite{video2shop}, Cheng \etal make a novel attempt to match clothes appeared in videos to the same items in online shops by cross-domain visual feature matching. In particular, a clothing detector is used to localize regions of interest in a given video. Given the difficulty in training good detectors for a wide range of products, their detector-based solution is not directly applicable in the current context that has tens of thousands of diverse products.
Moreover, as \cite{video2shop} is purely content-based, no multimodal representation learning is involved.} 
In \cite{real20m}, Chen \etal propose QCD for bidirectional retrieval between the product-page (P) and short-video (S) domains. In particular, QCD assumes each visual example, either a product image in the P domain or a short video in the S domain, is associated with a user query obtained based on click-through records. By cross-attention based fusion of the textual feature of the query and the visual feature extracted from the visual example, a multimodal product representation (MmPR) is obtained. Since user queries are relatively short, no text enhancement is involved in QCD. Moreover, for videos lacking click-through in particular those related to newly listed products, no query will be available. In \cite{rice}, Yang \etal propose RICE for retrieving product pages for a given live-stream video. In particular, given a product image with a title and a video with ASR text, their cross-domain similarity is calculated by combining the video-image similarity and the ASR-title similarity.  No ASR text enhancement is conducted. In a contemporary work \cite{rope}, Bai \etal consider a more realistic yet more challenging scenario of three-domain CdPR. However, their proposed COPE method is fully visual based, without exploiting the text modality\footnote{After analyzing the official code of COPE, we find that the title channel of the product-page domain was also unused.}. We follow the challenging setting of COPE, and go one step further by exploiting the ASR text for obtaining MmPR.

\begin{table}[htbp]
\caption{\textbf{Current methods for Cross-domain Product Retrieval (CdPR)}. P: Product-page. S: Short-video. L: Live-stream. MmPR: Multimodal Product Representation.} 
\label{tab:re_cdpr}
\renewcommand{\arraystretch}{1.1}
\centering
\resizebox{\linewidth}{!}{
\begin{tabular}{llcccc}
\toprule
\textbf{Method} & \textbf{\makecell{Product \\domains}} & \textbf{CdPR tasks} & \textbf{Text input} & \textbf{MmPR?} & \textbf{\makecell{Text \\enhancement?}} \\
\midrule
\re{Video2Shop, CVPR17\cite{video2shop}} & P, S & S2P & -- & \xmark & \xmark \\
\rowcolor{gray!10}  QCD, MM23\cite{real20m} & P, S & P2S, S2P & user query & \cmark & \xmark \\
RICE, ICCV23\cite{rice} & P, L & L2P & ASR text & \xmark & \xmark \\
\rowcolor{gray!10} COPE, ICCV23\cite{rope} & P, S, L & \makecell{P2L, P2S, S2P,\\S2L, L2P, L2S} & -- & \xmark & \xmark \\
AMPere (\emph{this paper}) & P, S, L & \makecell{P2L, P2S, S2P,\\S2L, L2P, L2S} & ASR text & \cmark & \cmark \\
\bottomrule
\end{tabular}
}
\end{table}

\subsection{The Use of ASR Text for Video Analysis}

The ASR text has been frequently used in varied video tasks. In text-video retrieval, for instance, the ASR text is typically vectorized by word2vec and used together with other features \cite{liu2019CE,gabeur2020multi,shvetsova2022everything}. \re{For video hyperlinking, \cite{hyperlinking, cheng2017selection} extract bag-of-words feature from the ASR text.} For video decoration with sound effects, the ASR text is jointly exploited with video frames for multimodal video embedding \cite{aaai25-VDSFX}. The text is also used as relatively weak annotations for large-scale pre-training \cite{gan2023cnvid,tang2021decembert,howto100m}. All the above works use the ASR text as is albeit its noisy nature. For VideoQA, Zellers \etal \cite{zellers2021merlot} propose to add punctuation to the ASR with a sequence-to-sequence model that is trained to add punctuation to sentences / paragraphs from news articles. For generic text-video retrieval, Chen \etal \cite{table} employ KeyBERT \cite{keybert} to extract keywords from the ASR text. With the assumption that words most similar to a given document are good keywords for representing the document, KeyBERT calculates word-document similarities based on their BERT embeddings. 
In the E-commerce field, live streamers often spend much of their time chatting casually during product promotion sessions. The actual time dedicated to discussing the products is typically a small portion. As such, the ASR text transcribed from short or live-stream videos is too noisy to be handled by KeyBERT. To conquer the challenge, we make a novel attempt by harnessing an LLM for effective ASR text summarization.

\section{Proposed Method}\label{sec:met}
\subsection{Problem Formalization}

We formalize the problem of multimodal product representation learning (MmPRL) as follows. A specific product on an E-commerce platform is typically exhibited as data samples in the following three domains, \ie product-page (P), short-video (S), and live-stream (L). Let $x$ be a specific sample, with $x_p$, $x_s$ and $x_l$ be its three domain-specific instantiations. In particular, $x_p$ indicates an image of the given product, $x_s$ is a short video describing the product, whilst $x_l$ denotes a longer live-stream video with a network anchor prompting the product. In a multimodal scenario, the image $x_p$ is associated with a title $t_p$, while the two videos $x_s$ and $x_l$  are associated with ASR texts $t_s$ and $t_l$, respectively.
For the ease of consistent description, we re-use $x$ to indicate a given multimodal sample, with its domain identity and the associated text both temporally omitted.
The goal of MmPRL is to train a feature extraction network $\mathcal{F}$ that encodes the given sample into a $d$-dimensional embedding $e(x)$. Accordingly, CdPR boils down to calculating cosine similarities between $e(x_p)$, $e(x_s)$ and $e(x_l)$.

Our method is developed based on the state-of-the-art COPE \cite{rope}. So we first describe COPE briefly in Sec. \ref{ssec:cope}, followed by our LLM-based text summarizer that handles the noisy ASR text in Sec. \ref{ssec:preprocess}. Our multimodal extension of COPE is detailed in Sec. \ref{ssec:model}.

\subsection{COPE in a Nutshell} \label{ssec:cope}

To handle visual inputs from the three domains, the COPE network has three branches. Each branch consists of two modules, \ie a video encoder that converts the corresponding input $x$ into a $512$-dimensional video feature $v(x)$, followed by a linear layer to obtain a shorter $128$-d embedding $e(x)$ for cross-domain matching. 

COPE adopts the video encoder of X-CLIP \cite{ni2022expanding}, which has a novel cross-frame attention mechanism to capture the long-range temporal dependencies across frames. More specifically, given a video $x$ as a sequence of $n$ uniformly sampled frames $\{f_1, \ldots, f_n\}$, a ViT \cite{vit} based cross-frame communication transformer (CCT) is used to jointly extract a sequence of $512$-dimensional frame-level features $\{z_1, \ldots, z_n\}$. The frame features then go through a stack of four standard Transformer blocks followed by average pooling to produce the video-level feature. More formally, we have
\begin{equation} \label{eq:cope}
\resizebox{.9\hsize}{!}{$
\left. \begin{array}{ll}
 \{f_1, \ldots, f_n\} & \leftarrow \mbox{video-to-frames}(x, n=8), \\
 \{z_1, \ldots, z_n\} & \leftarrow \mbox{CCT}(\{f_1, \ldots, f_n\}),\\
 v(x) & \leftarrow \mbox{AvgPool}(\mbox{Transformers}(\{z_1, \ldots, z_n\})),\\
 e(x) & \leftarrow \mbox{Linear}_{512 \times 128}(v(x)).
       \end{array} \right. $}
\end{equation}

In COPE, the parameters of the video encoder are shared across the three branches, while the parameters of the linear layer in Eq. \ref{eq:cope} are branch-specific. To handle the image input of the product-page domain, COPE simply duplicates the image $n$ times to form a sequence of pseudo frames.

It is worth noting that although the dataset released by the COPE paper is multimodal with ASR texts available, the original paper has not utilized the text modality because ``\emph{the excessive noise information in raw ASR text can negatively impact the final presentations}'' \cite{rope}, making COPE exclusively visual-based. With a simple LLM-based text summarizer, we will make the ASR text usable. 

\subsection{LLM-based ASR Text Summarization} \label{ssec:preprocess}

Given the verbose ASR text obtained from a video of a specific product, say an electric shaver, we aim to extract two key pieces of product-specific information. That is, the name of the product (\eg \emph{Space UFO shaped shaver}) and major features of the product (\eg \emph{high quality blade material, fast and clean shaving, \ldots}). To that end, we turn a \re{generic} LLM into an ASR text summarizer with the popular In-Context Learning (ICL) technique \cite{icl}, \re{which requires no model re-parameterization nor fine-tuning}. In essence, ICT performs a test-time adaptation of an LLM by instructing the model with a few demonstration examples. See Table \ref{tab:llm_prompt} for our demonstrations and our project website\footnote{\url{https://rucmm.github.io/mmrl4cdpr/}} for source code.

\begin{figure*}[htbp]
    \centering
    \begin{minipage}[c]{0.5\linewidth}
        \centering
        \includegraphics[width=\textwidth]{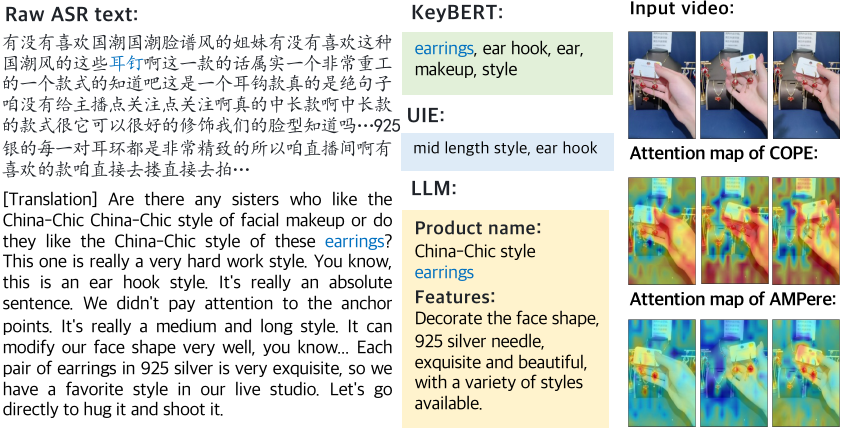}
        \centerline{(a) Product: \textit{earrings}}
    \end{minipage}
    \hspace{0.8em}
    \begin{minipage}[c]{0.45\linewidth}
        \centering
        \includegraphics[width=\textwidth]{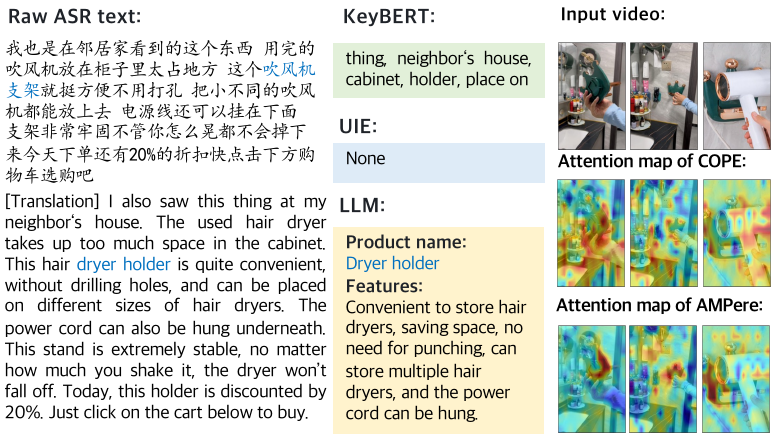}
        \centerline{(b) Product: \textit{dryer holder}}
    \end{minipage}
    \vspace{2mm}

    \begin{minipage}[c]{0.5\linewidth}
        \centering
        \includegraphics[width=\textwidth]{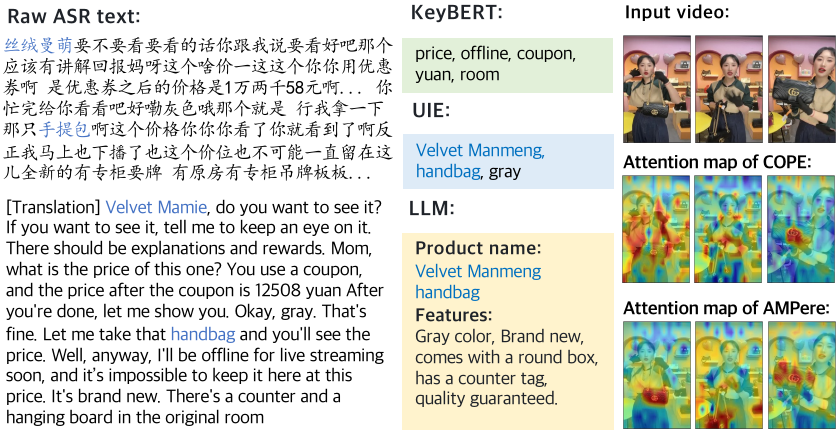}
        \centerline{(c) Product: \textit{sponge mop}}
    \end{minipage}
    \hspace{0.8em}
    \begin{minipage}[c]{0.45\linewidth}
        \centering
        \includegraphics[width=\textwidth]{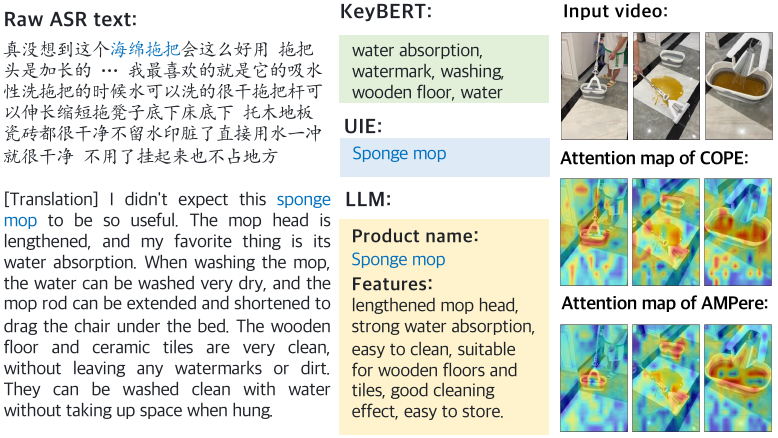}
        \centerline{(d) Product: \textit{velet manmeng handbag}}
    \end{minipage}
    \caption{\textbf{Qualitative results of ASR text summarization}. We compare with KeyBERT \cite{keybert}, a BERT embedding based keyword extraction tool, and UIE \cite{uie}, a competitive open-source model for information extraction. Text is shown partially due to space limit. The attention map per frame is obtained with the CLS token as queries (Q) and the patch tokens as keys (K). Since the CLS token is used as the frame-level feature for subsequent video feature extraction, the attention map shows the region the model is focusing on, for instance, the small hot spots corresponding to earrings in Fig. \ref{fig:asr_clean}a. Best viewed digitally.}
    \label{fig:asr_clean}
\end{figure*}

As the ROPE dataset \cite{rope} used in this study was gathered from a Chinese E-commerce platform with text in Chinese, a Chinese LLM is needed. Specifically, we adopt a 4bits-quantized version of the 13B-parameter Baichuan Chat model\footnote{\href{https://huggingface.co/baichuan-inc/Baichuan2-13B-Chat}{https://huggingface.co/baichuan-inc/Baichuan2-13B-Chat}}, a leading LLM on the CLiB Chinese LLM benchmark\footnote{\href{https://github.com/jeinlee1991/chinese-llm-benchmark}{https://github.com/jeinlee1991/chinese-llm-benchmark}}. This version strikes a good balance between model effectiveness and inference cost. See Fig. \ref{fig:asr_clean} for qualitative results of our LLM-based ASR text summarization. 

\re{\textbf{Handling missing ASR / LLM failure}. Note that for around 1\% of the ROPE videos, no ASR text is available, see Table \ref{tab:dataset}. For such ASR-missing cases, we simply fill out the missing value with a dummy string (``\textbackslash{N}\textbackslash{N}''  specified in ROPE). Meanwhile, for videos lacking product-related narration, \eg the audio channel is purely background music, the LLM may fail to product meaningful text, yielding output like ``Product name: Unknown; Features: Unknown''. We use the output \emph{as is} for multimodal representation learning.}

\subsection{Adding Summarized ASR to COPE} \label{ssec:model}

Our multimodal extension of COPE  is illustrated in Fig. \ref{fig:framework}. In order to accept the text modality, per branch we add a six-layer Chinese RoBERTa \cite{liu2019roberta} as a text encoder. Given an input text $t$ as a sequence of $m$ tokens $\{w_1,\ldots, w_m\}$, the text encoder yields $m$ $768$-d token-level features $\{y_1, \ldots, y_m\}$ plus a CLS-token feature $y_0$. 

For multimodal fusion, the visual and textual features have to be aligned beforehand. We use a linear layer to transform the textual features to a new sequence of 512-d features $\{\hat{y}_0, \hat{y}_1, \ldots, \hat{y}_m\}$. In a similar manner, we obtain a sequence of $n+1$ transformed visual features as $\{\hat{v}(x), \hat{z}_1, \ldots, \hat{z}_n\}$. The visual and textual feature sequences, concatenated as a $(m+n+2)\times 512$ multimodal tensor, are fed into an array of four Transformer blocks for intra-modality and inter-modality feature interaction and updating. We preserve the video feature and the CLS-token feature updated by the Transformers, denoted as $\bar{v}(x)$ and $\bar{y}_0$, respectively. The final multimodal embedding $e(x,t)$ is obtained by summing up $\bar{v}(x)$ and $\bar{y}_0$ followed by a linear layer. More formally, we express the above multimodal fusion process as follows:
\begin{equation} \label{eq:mm}
\resizebox{.9\hsize}{!}{$
\left. \begin{array}{ll}
 \{w_1,\ldots, w_m\} & \leftarrow \mbox{text-to-tokens}(t, m=32), \\
 \{y_0, y_1, \ldots, y_m\} & \leftarrow \mbox{RoBERTa}(\{w_1,\ldots, w_m\}), \\
  \{\hat{y}_0, \hat{y}_1, \ldots, \hat{y}_m\} & \leftarrow \mbox{Linear}_{768 \times 512}(\{y_0, y_1, \ldots, y_m\}), \\
  \{\hat{v}(x), \hat{z}_1, \ldots, \hat{z}_n\} & \leftarrow \mbox{Linear}_{512 \times 512}(\{v(x), z_1, \ldots, z_n\}), \\
 \{\bar{v}(x), \ldots, \bar{z}_n, \bar{y}_0, \ldots, \bar{y}_m\} & \leftarrow \mbox{Trans}(\{\hat{v}(x), \ldots, \hat{z}_n, \hat{y}_0, \ldots, \hat{y}_m\}) \\
 e(x,t) & \leftarrow \mbox{Linear}_{512 \times 128}(\bar{v}(x) + \bar{y}_0).
       \end{array} \right.$}
\end{equation}
Due to the use of the text encoder and the fusion modules, \texttt{AMPere} has more parameters than COPE (207M \emph{vs}. 134M). \re{By executing Eq. \ref{eq:mm} per branch, we obtain multimodal embeddings $e(x_p,t_p)$, $e(x_s,t_s)$, $e(x_l,t_l)$ for the P, S and L domains, respectively. Next, we describe how the three branches are jointly trained to ensure cross-domain embedding alignment.}

\subsection{Cross-domain Embedding Alignment}

\re{
Following COPE, we train our three-branch network by minimizing a combined loss which consists of three inter-domain \emph{pairwise} contrastive losses and three intra-domain \emph{pointwise} classification losses. In particular, given two specific domains, say P and S, an inter-domain contrastive loss is computed by treating $e(x_p,t_p)$ and $e(x_s,t_s)$ of the same product as a positive pair, with their counterparts from different products as negative pairs. Meanwhile, for each domain a classification loss is calculated by predicting the product label of the current instance based on its embedding $e(x,t)$. By jointly optimizing these six losses, multimodal embeddings of instances from different domains are \emph{product-wise} aligned.}

For each trainable layer in the network, its parameters are shared across the three branches, \re{\ie a single set of model parameters is used}. We empirically find that such a simple parameter-sharing strategy is better than alternatives such as training the last linear layer in a domain-specific manner. 
Except for the LLM, our network is end-to-end trained. Once trained, the network is used to produce multimodal embeddings $e(x_p,t_p)$, $e(x_s,t_s)$, $e(x_l,t_l)$ for bidirectional CdPR between the three domains.

\section{Experiments}\label{sec:exp}
\subsection{Experimental Setup}

\textbf{Dataset}. We adopt ROPE \cite{rope}, a public dataset released by Kuaishou (Kwai) for large-scale CdPR. With 3.1M product pages, 5.9M short videos, and 3.5M live streams for over 189K products, ROPE is currently the largest multi-domain multimodal product dataset of its kind. Tab. \ref{tab:dataset} shows data statistics of ROPE. We follow the official data split, 98.7\% for training and 1.3\% for testing.

\begin{table}[htbp]
\centering
\caption{\textbf{Statistics of the ROPE dataset we use}. Text length is the amount of Chinese characters per sample. }
\label{tab:dataset}
\resizebox{0.9\linewidth}{!}{
\begin{tabular}{llrr}
\toprule
\textbf{Domain} & \textbf{Statistics} & \textbf{Training set} & \textbf{Test set} \\ 
\midrule
\multirow{2}{*}{\emph{Product-page}} & \#Images & 3.0M & 31.4K \\
& Average text length & $35.0$ & $21.5$ \\
\midrule
\multirow{7}{*}{\emph{Short-video}} & \#Videos &5.8M & 29.8K \\
& Average duration & $35.4$s & $31.8$s \\
\cmidrule(lr){2-4}
& \re{\emph{ASR missing}} & \re{0.5\%} & \re{1.2\%} \\
& \re{\emph{LLM No-output}} & \re{9.9\%} & \re{25.1\%} \\
\cmidrule(lr){2-4}
& Average text length: \\
& \emph{RAW ASR} & $244.1$ & $214.0$ \\
& \emph{UIE summarized} & $9.4$ & $7.6$ \\
& \emph{LLM summarized} & $59.6$ & $52.1$ \\
\midrule
\multirow{7}{*}{\emph{Live-stream}} & \#Videos & 3.5M & 31.0K \\
& Average duration & $112.2$s & $129.1$s \\
\cmidrule(lr){2-4}
& \re{\emph{ASR missing}} & \re{0.4\%} & \re{0.1\%} \\
& \re{\emph{LLM No-output}} & \re{3.6\%} & \re{3.3\%} \\
\cmidrule(lr){2-4}
& Average text length: \\
& \emph{RAW ASR} & $543.6$ & $550.0$ \\
& \emph{UIE summarized} & $13.6$ & $9.8$ \\
& \emph{LLM summarized} & $62.6$ & $66.8$ \\
\bottomrule
\end{tabular} 
}
\end{table}

\textbf{Tasks}. Following \cite{rope}, we evaluate  the proposed method for the following six CdPR tasks. That is, P2S, P2L, S2P, S2L, L2P, and L2S, where P, S, and L indicate the product-page, short-video, and live-stream domains, respectively. 

\textbf{Evaluation criteria}. 
Per task, we report standard rank-based metrics: R1, R5 and R10.
\ie{recall at top k (k=1, 5, 10)}. 
For overall comparison, we use mean R1 of all tasks (mR1).
\re{In addition, we report Mean Reciporal Rank (MRR) and Normalized Discounted Cumulative Gain at 10 (NDCG$_{10}$).}

\textbf{Implementation details}. As mentioned earlier, our solution is developed on top of COPE \cite{rope}. So for fair comparison, we follow the implementation of this baseline wherever applicable: the video encoder is initialized by X-CLIP-B/16 \cite{ni2022expanding} pretrained on Kinetics-600 \cite{kay2017kinetics}, the number of frames $n$ is set to 8. The instance-level contrastive loss \cite{oord2018representation} and the product-level classification loss \cite{an2022killing} is jointly minimized by the AdamW optimizer \cite{LoshchilovH19}. The maximum length of text tokens is empirically set to 32. The batch size is 128.  The learning rate is first warmed up by 4 epochs to 5e-5, 1e-5, 5e-5, and 5e-3 for the text encoder, the video encoder, the fusion transformers, and the other layers, respectively, and then adjusted with a cosine annealing decay schedule. For all models, we obtain their stable checkpoints at the 70th training epoch. Experiments are conducted with PyTorch on 16 NVIDIA Tesla V100 GPUs.

\subsection{Comparison with Baseline Methods}

\textbf{Baselines}. Based on our analysis of current methods for CdPR, \emph{c.f.} Table \ref{tab:re_cdpr},  COPE is a natural baseline to \texttt{AMPere}. Moreover, we compare with CN-CLIP \cite{cnclip}. This Chinese variant of CLIP is a widely accepted baseline for both generic text-to-video retrieval \cite{chinaopen} and CdPR \cite{rope,rice,real20m}. Per video, we obtain its CN-CLIP feature by mean pooling over the 8 frame-level features.  To check if \re{raw ASR text or} LLM-summarized text is also beneficial for CN-CLIP, we implement three variants:
\begin{itemize}
    \item \textbf{CN-CLIP-m0}: \re{Multimodal feature is obtained by averaging the video-level feature and the CN-CLIP feature of associated raw ASR text,} under zero-shot setting.
    \item \textbf{CN-CLIP-m1}: \re{Similar to CN-CLIP-m0}, yet using LLM-summarized ASR text instead of raw ASR text. 
    \item \textbf{CN-CLIP-m1-finetune}: CN-CLIP-m1, with CN-CLIP finetuned on ROPE with LLM-summarized ASR text.
\end{itemize}

We further compare with QCD \cite{real20m}, \re{another multimodal method,} under two settings: 
\begin{itemize}
    \item \textbf{QCD}: QCD trained on ROPE with raw ASR text.
    \item \textbf{QCD+}: QCD trained on ROPE with LLM-summarized ASR text.
\end{itemize}

In addition, we compare with mPLUG-Video \cite{mplug-video}, a Chinese large multimodal model. To check if such a general-purpose model can be directly used for MmPRL and if it can be improved by fine-tuning, we try mPLUG-Video in the following two settings. Both use LLM-summarized ASR text.  
\begin{itemize}
    \item \textbf{mPLUG-Video-0}: Per video-text pair, the last token of the language decoder of mPLUG-Video\footnote{\url{https://huggingface.co/MAGAer13/mplug-youku-bloomz-7b}}, with embedding size of 4,096, is used as multimodal feature.
    \item \textbf{mPLUG-Video-\emph{lb}}: mPLUG-Video in a linear probing (\emph{lb}) mode, where the previous multimodal feature is fed to a 4,096$\times$128 linear layer, trained on ROPE.
\end{itemize}

\re{Note that except for COPE and CN-CLIP, all the other baselines have utilized ASR text (or its clean version) for multimodal representation.}

\begin{figure}[!ht]
    \begin{minipage}[c]{\linewidth}
        \centering
        \includegraphics[width=1.0\textwidth]{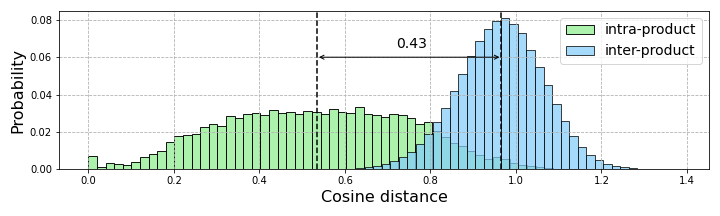}
        \centerline{(a) COPE}
    \end{minipage}
    \vspace{2mm}
    
    \begin{minipage}[c]{\linewidth}
        \centering
        \includegraphics[width=1.0\textwidth]{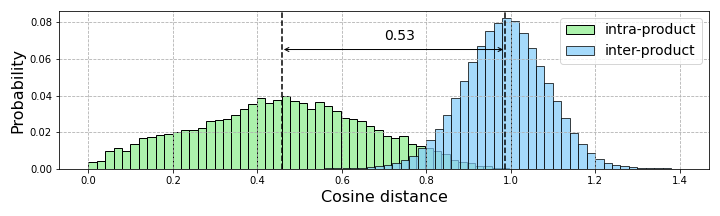}
        \centerline{(b) \texttt{AMPere}}
    \end{minipage}
    \caption{\textbf{Distribution of intra-product and inter-product distances} in the (a) COPE and (b) \texttt{AMPere} spaces, respectively.}
    \label{fig:tsne_and_hist}
\end{figure}

\begin{figure}[htbp]
    \centering
    \begin{minipage}[c]{0.49\linewidth}
        \centering
        \includegraphics[width=0.95\textwidth]{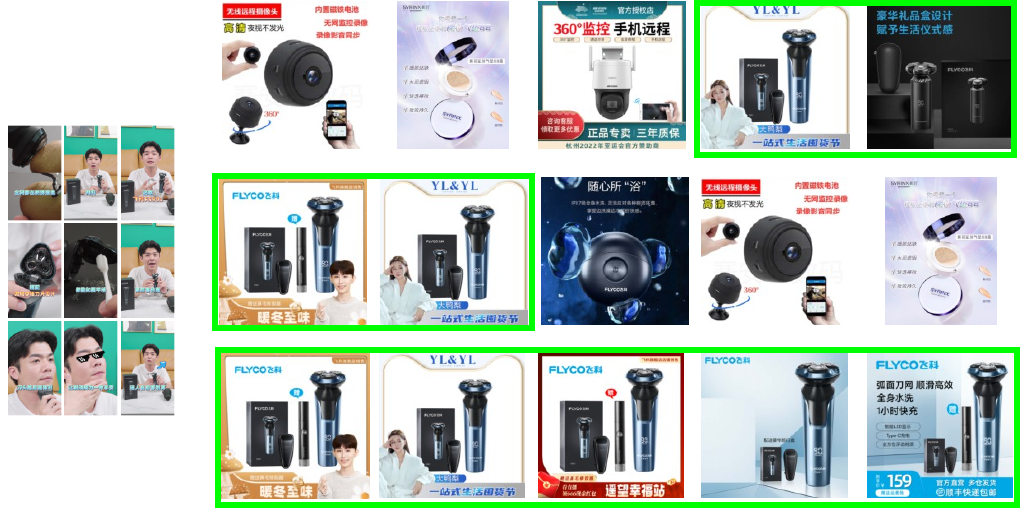}
        \centerline{(a) S2P (\textit{electric shaver})}
    \end{minipage}
    \begin{minipage}[c]{0.49\linewidth}
        \centering
        \includegraphics[width=0.97\textwidth]{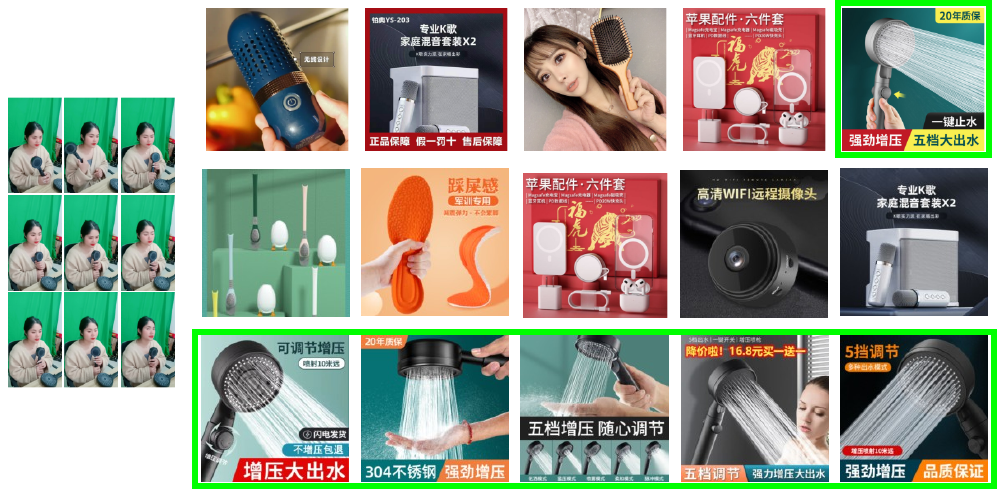}
        \centerline{(b) L2P (\textit{shower head})}
    \end{minipage}
    \vspace{2mm}
    
    \begin{minipage}[c]{0.49\linewidth}
        \centering
        \includegraphics[width=0.95\textwidth]{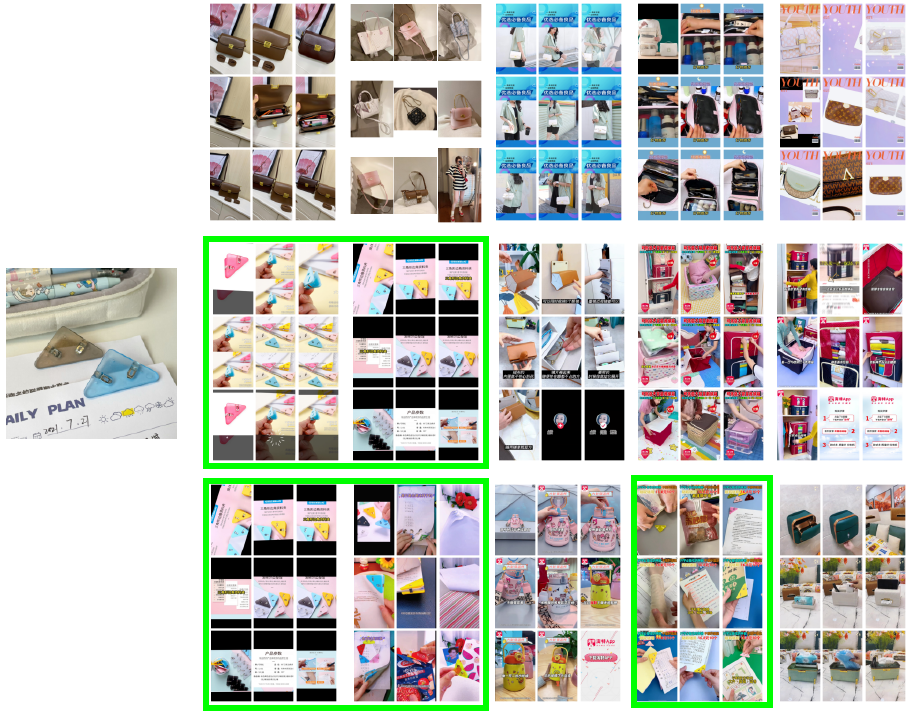}
        \centerline{(c) P2S (\textit{triangle clamp})}
    \end{minipage}
    \begin{minipage}[c]{0.49\linewidth}
        \centering
        \includegraphics[width=0.95\textwidth]{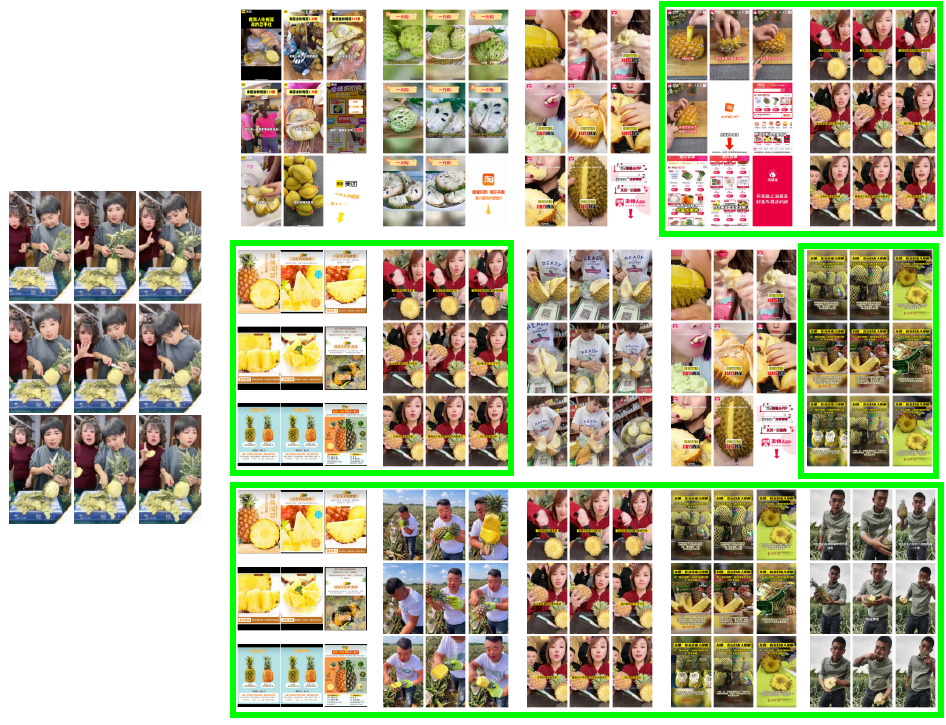}
        \centerline{(d) L2S (\textit{pineapple})}
    \end{minipage}
    
    \vspace{2mm}
    \begin{minipage}[c]{0.49\linewidth}
        \centering
        \includegraphics[width=0.95\textwidth]{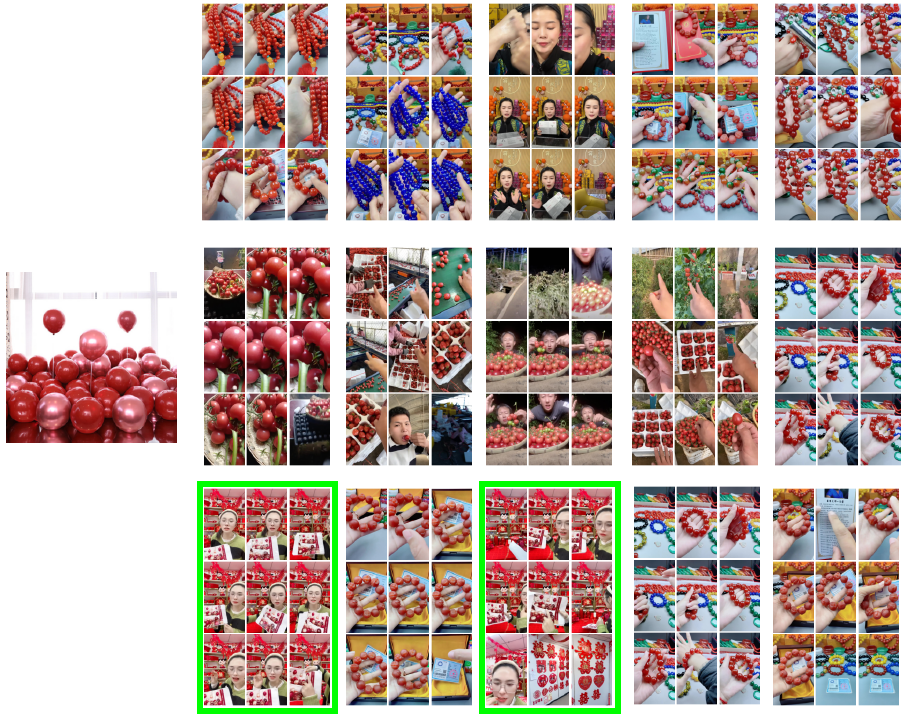}
        \centerline{(e) P2L (\textit{wedding balloon})}
    \end{minipage}              
    \begin{minipage}[c]{0.49\linewidth}
        \centering
        \includegraphics[width=0.95\textwidth]{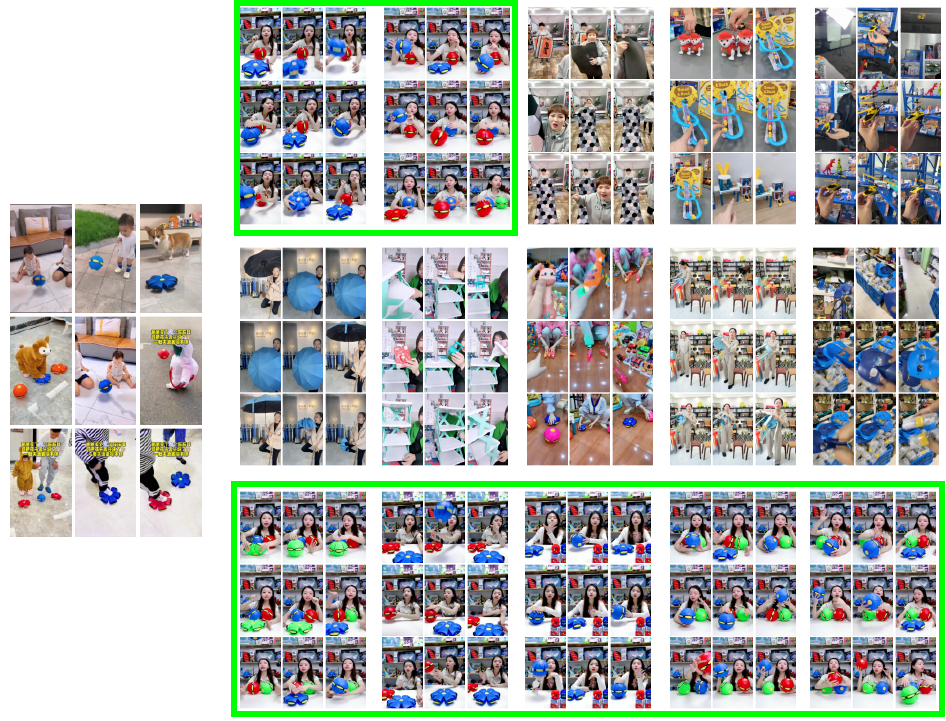}
        \centerline{(f) S2L (\textit{scalable football})}
    \end{minipage}

    \caption{\textbf{Hand-picked qualitative results}. Per sub-figure, from top to bottom, are top-5 results retrieved by COPE, \texttt{AMPere} with raw ASR, and \texttt{AMPere}, respectively. Correct results are shown in green. Best viewed on screen.}
    \label{fig:cdpr-results}
\end{figure}

\textbf{Results}. The performance of the different models is shown in Table \ref{tab:all_baseline}. Given COPE as the baseline, \texttt{AMPere} improves mR1 from 56.4 to 62.1, \ie a relative improvement of 10.5\%. \re{Our model is also the best in terms of MRR and NDCG$_{10}$, see Table \ref{tab:mrr} and \ref{tab:ndcg}.} The attention maps in Fig. \ref{fig:asr_clean} reveal that \texttt{AMPere} is more focused on product regions. Compared to COPE, \texttt{AMPere} has a smaller intra-product distance (0.53 \emph{vs}. 0.46) and a larger inter-product distance (0.96 \emph{vs}. 0.99), see Fig. \ref{fig:tsne_and_hist}, showing that the latter yields a more discriminative space for cross-domain product representation.  See \re{Fig. \ref{fig:failed_cases}} and Fig. \ref{fig:cdpr-results} for qualitative results. The superiority of \texttt{AMPere} over COPE is thus verified.

\begin{table*}[htbp!]
\centering
\caption{\textbf{Comparison with SOTA}.}
\label{tab:all_baseline}
\setlength{\tabcolsep}{3pt}
\renewcommand{\arraystretch}{1}
\resizebox{\linewidth}{!}{
\begin{tabular}{lrrr|rrr|rrr|rrr|rrr|rrr|r@{\hspace{3pt}}l}
\toprule

\multirow{2}{*}{\textbf{Model}} & 
\multicolumn{3}{c}{\textbf{P2S}} &\multicolumn{3}{c}{\textbf{P2L}} & \multicolumn{3}{c}{\textbf{S2P}} & \multicolumn{3}{c}{\textbf{S2L}} & \multicolumn{3}{c}{\textbf{L2P}} &
\multicolumn{3}{c}{\textbf{L2S}} &
\multirow{2}{*}{\textbf{mR1}} \\ 

\cmidrule(r){2-4} \cmidrule(r){5-7} \cmidrule(r){8-10} \cmidrule(r){11-13} \cmidrule(r){14-16} \cmidrule(r){17-19}
& R1 & R5 & R10 & R1 & R5 & R10 & R1 & R5 & R10 & R1 & R5 & R10 & R1 & R5 & R10 & R1 & R5 & R10 \\ 
\midrule
\emph{Visual-only}: & \\
\rowcolor{gray!20}  COPE  & 82.6 & 94.9 & 97.5 & 54.1 & 71.1 & 77.1 & 65.2 & 76.6 & 82.0 & 46.0 & 63.6 & 70.6 & 42.3 & 56.5 & 63.7 & 48.3 & 67.2 & 74.7 & 56.4 & \\
\emph{Cross-modal}: & \\
CN-CLIP  & 58.8 & 78.6 & 85.5 & 32.6 & 46.1 & 52.9 & 35.7 & 53.1 & 60.5 & 26.0 & 39.0 & 50.0 & 18.9 & 33.4 & 40.5 & 28.4 & 46.6 & 54.5 & 33.4 & (\drop{40.1\% $\downarrow$}) \\
 [3pt]
\emph{Multimodal}: \\
CN-CLIP-m0 & 55.5 & 82.4 & 90.1 & 38.3 & 62.7 & 71.3 & 19.7 & 34.0 & 41.2 & 22.8 & 39.6 & 48.7 & 9.3  & 17.1 & 22.0 & 15.4 & 33.4 & 43.4 & 26.8 & (\drop{52.5\% $\downarrow$}) \\ 
CN-CLIP-m1  & 63.2 & 85.7 & 92.4 & 48.5 & 70.0 & 78.3 & 34.6 & 52.3 & 59.1 & 30.1 & 46.6 & 54.1 & 25.5 & 42.1 & 49.9 & 31.8 & 52.4 & 61.6 & 39.0 & (\drop{30.9\% $\downarrow$}) \\
{CN-CLIP-m1-finetune} & 76.1 & 92.3 & 96.0 & 52.5 & 71.6 & 79.0 & 64.2 & 79.0 & 64.6 & 45.5 & 64.3 & 72.3 & 42.1 & 58.8 & 66.7 & 44.5 & 65.9 & 73.7 & 54.2 & (\drop{3.9\% $\downarrow$})\\
{QCD} & 69.3 & 89.2 & 94.5 & 52.3 & 74.4 & 82.5 & 63.7 & 79.2 & 84.4 & 47.3 & 67.7 & 75.5 & 42.1 & 59.2 & 66.3 & 43.0 & 65.3 & 73.5 & 52.9 & (\drop{6.2\% $\downarrow$}) \\
{QCD+} & 75.5 & 91.9 & 96.3 & 60.7 & 78.1 & 84.4 & 69.1 & 80.5 & 86.1 & 53.4 & 69.2 & 76.3 & 47.4 & 63.0 & 70.6 & 49.6 & 67.2 & 75.5 & 59.3 & (\gain{5.1\% $\uparrow$}) \\
mPLUG-Video-0 & 0.1 & 0.5 & 0.8 & 0.4 & 1.0 & 1.2 & 0.2 & 0.5 & 0.6 & 0.4 & 1.0 & 1.1 & 0.1 & 0.4 & 0.5 & 0.1 & 0.4 & 0.6 & 0.2 & (\drop{99.6\% $\downarrow$}) \\
mPLUG-Video-\emph{lb} & 50.7 & 76.7 & 86.1 & 38.3 & 63.2 & 73.1 & 40.7 & 62.5 & 71.6 & 29.8 & 51.1 & 62.2 & 26.2 & 46.9 & 58.4 & 27.5 & 50.3 & 60.9 & 35.5 & (\drop{37.1\% $\downarrow$}) \\ [3pt]

Proposed \emph{AMPere} & \textbf{84.5} & \textbf{95.9} & \textbf{98.3} & \textbf{65.0} & \textbf{81.4} & \textbf{86.5} & \textbf{71.2} & \textbf{84.3} & \textbf{88.6} & \textbf{54.7} & \textbf{71.3} & \textbf{77.5} & \textbf{47.8} & \textbf{64.6} & \textbf{72.2} & \textbf{50.4} & \textbf{68.9}  & \textbf{76.2} & \textbf{62.3} & (\gain{10.5\% $\uparrow$}) \\
\bottomrule
\end{tabular}
}
\end{table*}

\begin{table}[htbp!]
\centering
\caption{\scalebox{0.93}{\re{\textbf{Comparison with SOTA}. Metric: MRR.}}}
\label{tab:mrr}
\setlength{\tabcolsep}{3pt} 
\renewcommand{\arraystretch}{1} 
\resizebox{\linewidth}{!}{
\begin{tabular}{lrrrrrr@{\hspace{3pt}}l}
\toprule

\textbf{Model} & P2S & P2L & S2P & S2L & L2P & L2S & Mean \\ 
\midrule
 COPE & 0.892 & 0.639 & 0.761 & 0.596 & 0.503 & 0.580 & 0.662 \\
CN-CLIP & 0.678 & 0.397 & 0.440 & 0.326 & 0.262 & 0.372 & 0.412 \\
 [3pt]
CN-CLIP-m0 & 0.673 & 0.494 & 0.270 & 0.313 & 0.137 & 0.243 & 0.355 \\ 
CN-CLIP-m1  & 0.731 & 0.585& 0.430 & 0.383 & 0.336 & 0.415 & 0.480 \\
{CN-CLIP-m1-finetune} & 0.831 & 0.615 & 0.710 & 0.544 & 0.504 & 0.545 & 0.625 \\
{QCD} & 0.782 & 0.599 & 0.679 & 0.569 & 0.496 & 0.510 & 0.606 \\
{QCD+} & 0.830 & 0.664 & 0.763 & 0.617 & 0.541 & 0.585 & 0.667 \\
mPLUG-Video-0 & 0.005 & 0.005 & 0.005 & 0.006 & 0.003 & 0.005 & 0.005 \\
mPLUG-Video-\emph{lb} & 0.624 & 0.500 & 0.508 & 0.402 & 0.364 & 0.385 & 0.464 \\ 

\texttt{AMPere} & \textbf{0.895} & \textbf{0.726} & \textbf{0.767} & \textbf{0.628} &  \textbf{0.558} & \textbf{0.595} & \textbf{0.695} \\
\bottomrule
\end{tabular}
}
\end{table}

\begin{table}[htbp!]
\centering
\caption{\re{\textbf{Comparison with SOTA}. Metric: NDCG$_{10}$.}}
\label{tab:ndcg}
\setlength{\tabcolsep}{3pt} 
\renewcommand{\arraystretch}{1} 
\resizebox{\linewidth}{!}{
\begin{tabular}{lrrrrrr@{\hspace{3pt}}l}
\toprule

\textbf{Model} & P2S & P2L & S2P & S2L & L2P & L2S & Mean \\ 
\midrule

COPE & 0.763 & 0.503 & 0.680 & 0.470 & 0.399 & 0.437 & 0.542 \\
CN-CLIP & 0.398 & 0.285 & 0.319 & 0.233 & 0.167 & 0.208 & 0.268 \\
 [3pt]
CN-CLIP-m0 & 0.428 & 0.322 & 0.168 & 0.190 & 0.074 & 0.113 & 0.216 \\ 
CN-CLIP-m1 & 0.516 & 0.419 & 0.318 & 0.258 & 0.231 & 0.246 & 0.332 \\
{CN-CLIP-m1-finetune} & 0.655 & 0.474 & 0.607 & 0.419 & 0.396 & 0.384 & 0.489 \\
{QCD} & 0.577 & 0.486 & 0.632 & 0.427 & 0.365 & 0.328 & 0.469  \\
{QCD+} & 0.635 & 0.565 & 0.687 & 0.488 & 0.457 & 0.446 & 0.546 \\
mPLUG-Video-0 & 0.003 & 0.001 & 0.002 & 0.002 & 0.001 & 0.001 & 0.002  \\
mPLUG-Video-\emph{lb} & 0.389 & 0.328 & 0.362 & 0.257 & 0.249 & 0.221 & 0.301 \\ 

\texttt{AMPere} & \textbf{0.778} & \textbf{0.596} & \textbf{0.699} & \textbf{0.500} & \textbf{0.475} & \textbf{0.452} & \textbf{0.583} \\
\bottomrule
\end{tabular}
}
\end{table}

The clearly lower performance of the CN-CLIP and mPLUG-Video series to COPE suggests the inapplicability of the generic model for the E-commerce field. The better performance of CN-CLIP-m1 than CN-CLIP-m0 (39.0 \emph{versus} 26.8 in mR1) and QCD+ than QCD (59.3 \emph{versus} 52.9 in mR1) indicates that CN-CLIP and QCD also benefit from our ASR text summarization. 

\subsection{Understanding \texttt{AMPere}} \label{ssec:ablation}

In order to understand how \texttt{AMPere} works and particularly what elements contribute to its superior performance, we conduct a number of ablation studies as follows. For efficiency consideration, we randomly select one-tenth of the training data, with the test set unchanged. So the reported performance of a specific setup will be lower than its counterpart trained on the full-scale dataset.

\textbf{The role of the ASR text}. We begin by investigating the impact of ASR text on the CdPR performance. The following four  choices of the ASR-text input are compared:  
\begin{itemize}
    \item Raw ASR: The raw ASR text is directly used.
    \item  KeyBERT: We use KeyBERT \cite{keybert} to extract five keywords per ASR text. 
    \item UIE: We adopt the Universal Information Extraction (UIE) model \cite{uie}, a leading open-source tool\footnote{\href{https://huggingface.co/PaddlePaddle/uie-base}{https://huggingface.co/PaddlePaddle/uie-base}}, to extract product-related entities. To that end, the schema in UIE is empirically set to [`Brand', `Product category', `Product name', `Appearance', `Texture']. The UIE-summarized text is then used. 
    \item LLM-ProName: Using only the product-name field of the LLM-summarized text. 
    \item LLM-ProFeat: Using only the product-feature field. 
    \item LLM: Using both product name and features.
\end{itemize}
Per choice, we consider both the unimodal (\emph{without} the visual input) and the multimodal (\emph{with} the visual input) scenarios.

As shown in Table \ref{tab:ab_asr}, LLM performs the best in both scenarios. We observe that the LLM-summarized text is richer than its UIE counterpart, see Fig. \ref{fig:asr_clean}. This is also supported by the fact that the LLM-summarized text is on average much longer, see Table \ref{tab:dataset}. Given the uncontrolled nature of the ASR text, the LLM-based text summarizer may yield no result. However, the chance of no result is much lower than its UIE counterpart, see Table \ref{tab:asr_sum}.

Comparing the two fields, the higher CdPR performance of the product-name field than the product-feature field suggests that the former is more important.

Considering the multimodal scenario, the gain of adding the raw ASR  (0.7\%) or the UIE-summarized text (1.6\%) compared to the visual-only solution, \ie COPE, is marginal. By contrast, using the LLM-summarized text obtains a relative improvement of 5.8\%, see the last row of Table \ref{tab:ab_asr}. The result verifies the necessity of ASR text processing and the effectiveness of our ASR text summary. \re{The superior  performance of \texttt{AMPere} against the multimodal baseline of ``Visual + Raw ASR'' (60.2 \emph{vs}. 57.3 in mR1) allows us to attribute the improvement to the advantages of \texttt{AMPere} rather than the inclusion of ASR text.}

\begin{table}[htbp]
\caption{\textbf{Evaluating different usages of ASR text}.} 
\label{tab:ab_asr}
\setlength{\tabcolsep}{4pt}
\renewcommand{\arraystretch}{1}
\centering
\resizebox{\linewidth}{!}{
\begin{tabular}{lrrrrrrr@{\hspace{4.2pt}}l}
\toprule
\textbf{Text input} & P2S & P2L & S2P & S2L & L2P & L2V & mR1 & \\
\midrule
\multicolumn{8}{l}{Unimodal (\textit{without} visual input):}\\
\rowcolor{gray!20}
Raw ASR & 30.0 & 15.8 & 10.8 & 4.8  & 4.1  & 3.4  & 11.5 & \\
KeyBERT & 20.6 & 12.5 & 6.1 & 5.5 & 4.0 & 5.2 & 9.0 & (\drop{21.7\%} $\downarrow$) \\ 
UIE & 30.1 & 19.3 & 8.3  & 4.9  & 4.9  & 4.2  & 12.0 &(\gain{4.3\%} $\uparrow$) \\ [3pt]

LLM-ProFeat & 24.1 & 19.3 & 10.3 & 8.5  & 8.5  & 7.8  & 13.1 & (\gain{13.9\%} $\uparrow$) \\
LLM-ProName & 45.2 & 32.8 & 25.1 & 18.7 & 19.1 & 19.2 & 26.7 & (\gain{132.2\%} $\uparrow$) \\
LLM & 46.7 & 36.3 & 26.7 & 19.8 & 21.5 & 20.7 & 28.6 & (\gain{148.7\%} $\uparrow$) \\
\midrule
\multicolumn{8}{l}{Multimodal (\textit{with} visual input):}\\

\rowcolor{gray!20}
 -- & 81.7 & 54.0 & 66.6 & 49.5 & 41.7 & 47.7 & 56.9 & \\
Raw ASR & 81.7 & 55.4 & 67.3 & 49.3 & 42.1 & 47.7 & 57.3 & (\gain{0.7\%} $\uparrow$) \\
KeyBERT & 81.6 & 55.3 & 66.4 & 48.7 & 42.6 & 48.1 & 57.1 & (\gain{0.4\%} $\uparrow$) \\
UIE & 82.0 & 55.6 & 67.7 & 50.1 & 42.2 & 49.4 & 57.8 & (\gain{1.6\%} $\uparrow$) \\ [3pt]
LLM-ProFeat & 82.0 & 56.1 & 67.6 & 50.3 & 44.3 & 49.5 & 58.3 & (\gain{2.5\%} $\uparrow$)\\
LLM-ProName & 82.2 & 59.9 & 69.0 & 52.2 & 45.6 & 49.8 & 59.8 & (\gain{5.1\%} $\uparrow$) \\
LLM & \textbf{82.1} & \textbf{60.7} & \textbf{69.4} & \textbf{52.3} & \textbf{46.1} & \textbf{50.4} & \textbf{60.2} & (\gain{5.8\%} $\uparrow$) \\
\bottomrule
\end{tabular}
}
\end{table}

\medskip

\textbf{Evaluation of the three text summarizers}. In order to directly evaluate the output of the three ASR text summarizers, we build a ground-truthed dataset as follows. From the test set, we randomly selected 200 raw ASR text samples from the short-video domain and the live-stream domain, respectively. Recall that the text summarizers are designed to extract two product-specific fields, \ie \emph{product-name} and \emph{product-features}. For product name extraction, we manually identify the product name by jointly checking each text sample and its corresponding video. Note that some videos are merely associated with background music or casual chat. Eventually, we successfully identified product names for 133 text samples from the short-video domain and 163 text samples from the live-stream domain.  As Tab. \ref{tab:asr_sum} shows, our LLM-based text summarizer is clearly better than its KeyBERT and UIE counterparts.

\begin{table}[htbp]
\centering
\caption{\textbf{Accuracy of different ASR text summarizers for product-name extraction}.}
\label{tab:asr_sum}
\setlength{\tabcolsep}{2pt}
\renewcommand{\arraystretch}{1}
\resizebox{\linewidth}{!}{
\begin{tabular}{llrrr}
\toprule

\textbf{ASR text summarizer} &&  \textbf{Short-video domain} & \textbf{Live-stream domain} & \textbf{Overall} \\ 
\cmidrule{1-1} \cmidrule{3-5}
KeyBERT && 21.8\% & 18.9\%  & 20.3\% \\
UIE && 48.1\% &  36.2\% &  41.6\% \\
LLM && \textbf{93.2}\% &  \textbf{91.4}\% &  \textbf{92.2}\% \\
\bottomrule
\end{tabular}
}
\end{table}

The performance of product features (attributes) extraction is reported in Tab. \ref{tab:eval-product-features}. As a specific product is typically associated with multiple attributes, making it essentially a multi-label problem, we report both recall and accuracy. Our LLM-based text summarize 
also clearly outperforms KeyBERT and UIE in product features (attributes) extraction.

\begin{table}[htbp]
\centering
\caption{\textbf{Evaluating product-attribute extraction}.}
\label{tab:eval-product-features}
\setlength{\tabcolsep}{3pt}
\renewcommand{\arraystretch}{1}
\resizebox{\linewidth}{!}{
\begin{tabular}{@{}lcc|cc|cc@{}}
\toprule

\multirow{2}{*}{\textbf{\makecell{ASR text \\summarizer}}} & 
\multicolumn{2}{c}{\textbf{Short-video domain}} &\multicolumn{2}{c}{\textbf{Live-stream domain}} & \multicolumn{2}{c}{\textbf{Overall}} \\ 

\cmidrule(r){2-3} \cmidrule(r){4-5} \cmidrule(r){6-7}
& \emph{Recall} & \emph{Accuracy} & \emph{Recall} & \emph{Accuracy} & \emph{Recall} & \emph{Accuracy} \\ 
\midrule
KeyBERT & 22.3\% & 10.2\% & 18.2\% & 7.4\% & 20.3\% & 8.8\%  \\
UIE & 16.0\% & 69.2\% & 14.5\% & 61.2\% & 15.3\% & 65.2\% \\
LLM & 88.7\% & 91.0\% & 82.3\% & 91.3\% & 82.0\% & 91.2\% \\
\bottomrule
\end{tabular}
}
\end{table}

\medskip 

\textbf{Ablations of other LLMs}. We investigate other LLMs by directly using the \texttt{AMPere} model trained with Baichuan2-13B-Chat-4bits cleaned text to see how our model works when replacing the current LLM in summarizer by two smaller alternatives at the inference time, \emph{i.e.} Baichuan2-7B-Chat-4bits and Qwen2-1.5B-Instruct\footnote{\url{https://github.com/QwenLM/Qwen2}}. As Tab. \ref{tab:ab_llm} shows, at a relatively small loss of mR1 (60.2 $\rightarrow$ 58.4, 3.0\%), GPU footprint is reduced by 12.0\% (13.7GB $\rightarrow$ 12.0GB).

\begin{table}[hb]
\caption{\textbf{Ablations of other LLMs}.} 
\label{tab:ab_llm}
\setlength{\tabcolsep}{3pt}
\renewcommand{\arraystretch}{1.1}
\centering
\resizebox{\linewidth}{!}{
\begin{tabular}{lrrrrrrrrl}
\toprule
\textbf{LLM for inference} & GPU footprint & P2S & P2L & S2P & S2L & L2P & L2S & mR1 \\
\midrule 

\rowcolor{gray!20}
Baichuan2-13B-Chat-4bits & 13.7 GB & \textbf{82.1} & \textbf{60.7} & \textbf{69.4} & \textbf{52.3} & \textbf{46.1} & \textbf{50.4} & \textbf{60.2} \\
Qwen2-1.5B-Instruct & 12.8 GB & 81.1 & 58.4 & 67.7 & 50.8 & 45.0 & 49.1 & 58.7 \\
Baichuan2-7B-Chat-4bits & 12.0 GB & 80.8 & 58.4 & 67.7 & 50.5 & 44.0 & 48.8 & 58.4 \\
\bottomrule
\end{tabular}
}
\end{table}

\medskip

\textbf{Choices of the multimodal fusion layer}. The following alternatives are compared:
\begin{itemize}
    \item SUM: The final multimodal embedding $e(x,t)$ is obtained as  $\hat{v}(x) + \hat{y}_0$, with no Transformer used.
    \item CAT: $e(x,t)$ is derived by concatenating $\hat{v}(x)$ and $\hat{y}_0$.
    \item XA-\emph{t}-as-\emph{Q}: Cross attention based fusion \cite{albef}, with the textual features $\{\hat{y}_0, \hat{y}_1, \ldots, \hat{y}_m\}$ as $Q$ and the visual features $\{\hat{v}(x), \hat{z}_1, \ldots, \hat{z}_n\}$ as $K$ and $V$. Such a module is also used in \cite{real20m} for CdPR.
    \item XA-\emph{v}-as-\emph{Q}: Similar to XA-\emph{t}-as-\emph{Q}, but with the role of the textual / visual features switched.
    \item CoA: Co-attention from ViLBERT \cite{vilbert}.
\end{itemize}

As shown in Table \ref{tab:ab_fusion}, our current choice is on par with CoA, yet uses only half the number of parameters.

\begin{table}[htbp]
\caption{\scalebox{0.95}{\textbf{Evaluating different choices of multimodal fusion}}.} 
\label{tab:ab_fusion}
\setlength{\tabcolsep}{5pt}
\renewcommand{\arraystretch}{1}
\resizebox{\linewidth}{!}{
\begin{tabular}{lrrrrrrrl}
\toprule
\textbf{Fusion} & \#Params (M) & P2S & P2L & S2P & S2L & L2P & L2S & mR1 \\
\midrule
\rowcolor{gray!20}
Ours & 12.8 & 82.1 & \textbf{60.7} & 69.4 & 52.3 & \textbf{46.1} & \textbf{50.4} & \textbf{60.2}  \\
SUM & 0.0 & 83.2 & 56.5 & 69.2 & 51.6 & 44.3 & 49.3 & 59.0  (\drop{2.0\%} $\downarrow$) \\
CAT & 0.0 & 81.0 & 59.9 & 68.5 & 49.5 & 45.7 & 49.3 & 59.0 (\drop{2.0\%} $\downarrow$) \\
XA-\emph{t}-as-\emph{Q} & 12.3 & 81.7 & 60.0 & 67.6 & 50.7 & 44.9 & 48.3 & 58.9 (\drop{2.2\%} $\downarrow$) \\
XA-\emph{v}-as-\emph{Q} & 12.3 & 82.0 & 59.3 & 68.0 & 51.4 & 44.0 & 48.7 & 58.9 (\drop{2.2\%} $\downarrow$) \\
CoA & 24.6 & \textbf{82.3} & 60.6 & \textbf{70.2} & \textbf{52.5} & 44.9 & 50.0 & 60.1 (\drop{0.2\%} $\downarrow$) \\

\bottomrule
\end{tabular}
}
\end{table}

\medskip

\textbf{Parameter sharing or not}? The COPE paper recommends training the last linear layer of each branch in a domain-specific manner \cite{rope}, yet without verifying the recommendation. To resolve such uncertainty, we train the multimodal model with the parameters of the last linear layer shared or not shared, respectively. 

As Table \ref{tab:ab_share} shows, parameter sharing is better. The same conclusion is obtained when training the unimodal models, either visual or textual. Our interpretation is that when the parameters are not shared, the three branches tend to produce domain-specific embeddings, thus making cross-domain embedding alignment more challenging. So, in contrast to \cite{rope}, we suggest parameter sharing.

\begin{table}[htbp]
\caption{\textbf{Evaluating the parameter sharing strategy}.} 
\label{tab:ab_share}
\setlength{\tabcolsep}{4pt}
\renewcommand{\arraystretch}{1}
\resizebox{\linewidth}{!}{
\begin{tabular}{@{}lcrrrrrrl@{}}
\toprule
\textbf{Input} & Sharing? & P2S & P2L & S2P & S2L & L2P & L2S & mR1 \\
\midrule 
\multirow{2}{*}{multimodal} & \scalebox{0.75}{\usym{2613}} & 80.2 & 47.9 & 65.9 & 42.6 & 35.2 & 42.1 & 52.3 \\
 & \checkmark & 82.1 & 60.7 & 69.4 & 52.3 & 46.1 & 50.4 & 60.2 (\gain{15.1\%} $\uparrow$) \\
\midrule 
\multirow{2}{*}{visual} & \scalebox{0.75}{\usym{2613}} &82.6 & 48.7 & 66.2 & 45.0 & 37.2 & 43.6 & 53.9 \\
 & \checkmark & 81.7 & 54.0 & 66.6 & 49.5 & 41.7 & 47.7 & 56.9 (\gain{5.6\%} $\uparrow$) \\
\midrule 
\multirow{2}{*}{text} & \scalebox{0.75}{\usym{2613}} & 43.4 & 31.5 & 24.9 & 17.7 & 19.3 & 18.7 & 25.9 \\
 & \checkmark & 46.7 & 36.3 & 26.7 & 19.8 & 21.5 & 20.7 & 28.6 (\gain{10.4\%}$\uparrow$) \\
\bottomrule
\end{tabular}
}
\end{table}

\medskip

\textbf{Contribution per domain}. To reveal the influence of the individual domains on MmPRL, we separately exclude each domain from training. As Table \ref{tab:ab_domain} shows, the short-video domain is the most important, followed by the product-page domain and the live-stream domain. With manual inspection, we observe that in the ROPE dataset, the short-video domain has a considerable amount of live-stream style videos. Whether this phenomenon stems from biases in the dataset gathering process or properties inherent to the E-commerce platform where the dataset was sourced requires further investigation.
\begin{table}[htbp]
\caption{{\textbf{Influence of the individual domains when excluded from training}. The short-video domain is the most influential. The live-stream domain is the least important.}}
\label{tab:ab_domain}
\renewcommand{\arraystretch}{1} 
\centering
\resizebox{\linewidth}{!}{
\begin{tabular}{lrrrrrrl}
\toprule
\textbf{Domains} & P2S & P2L & S2P & S2L & L2P & L2S & mR1 \\
\midrule 
\rowcolor{gray!20}
Full & 82.1 & 60.7 & 69.4 & 52.3 & 46.1 & 50.4 & 60.2 \\
\emph{wo/} Live stream & 81.9 & 60.4 & 69.4 & 50.1 & 47.1 & 49.0 & 59.7 (\drop{0.8\%} $\downarrow$) \\
\emph{wo/} Product page & 80.4 & 55.0 & 66.2 & 49.7 & 43.6 & 49.7 & 57.4 (\drop{4.7\%} $\downarrow$) \\
\emph{wo/} Short video & 79.0 & 50.3 & 58.8 & 38.5 & 38.5 & 43.9 & 51.5 (\drop{14.5\%} $\downarrow$) \\
\bottomrule
\end{tabular}
}
\end{table}

\medskip

\re{\textbf{Robustness analysis}. In order to reveal how \texttt{AMPere} performs under varying ASR data quality, per ASR text we calculate its signal level as the ratio of the LLM-summarized text length to the raw ASR text length. A bigger ratio indicates a larger proportion of product-related information within the raw text, and thus a stronger signal level. Accordingly, we group test queries according to the signal level of their associated ASR text. As shown in Table \ref{tab:robustness}, \texttt{AMPere} consistently outperforms the baselines at varied ASR signal levels. Even in the extreme case of ZERO signal, \ie \emph{LLM no-output}, \texttt{AMPere} is also better than COPE (38.1 \emph{vs} 36.2 in R1). We attribute this result to the fact that with multimodal learning, the visual encoder of \texttt{AMPere} is stronger than its counterpart in COPE, see Fig. \ref{fig:asr_clean}. As such, even the test queries lack meaningful ASR text input, using the multimodal model remains beneficial.}

\begin{table}[htbp]
\centering
\caption{\re{\textbf{Performance under varying ASR data quality}. Query domain: L / Retrieval domain: P. Metric: R1.} }
\label{tab:robustness}
\setlength{\tabcolsep}{3pt}
\renewcommand{\arraystretch}{1}
\resizebox{\linewidth}{!}{
\begin{tabular}{l|r|r|r|r}
\toprule

\textbf{ASR Signal Level} & \textbf{Percentage} & \textbf{Visual only} & \textbf{Visual + Raw ASR} & \textbf{AMPere} \\

\midrule
\rowcolor{gray!20}
0 (\emph{LLM No-output}) & 3.4\% & 36.2 & 33.4 & \textbf{38.1} \\
$(0, 0.05)$ & 10.4\% & 35.5 & 37.3 & \textbf{39.2} \\
$[0.05, 0.1)$ & 48.1\% & 43.7 & 43.3 & \textbf{48.2} \\
$[0.1, 0.15)$ & 23.1\% & 43.6 & 44.3 & \textbf{47.3} \\
$[0.15, 0.2)$ & 8.2\% & 37.6 & 40.6 & \textbf{46.5} \\
$[0.2, 0.25)$ & 3.7\% & 39.7 & 39.1 & \textbf{42.8} \\
$>=0.25$ & 3.1\% & 38.9 & 40.7 & \textbf{41.3} \\
\bottomrule
\end{tabular}
}
\end{table}

\medskip

\begin{figure}[!htb]
    \begin{minipage}[c]{\linewidth}
        \centering
        \includegraphics[width=1.0\textwidth]{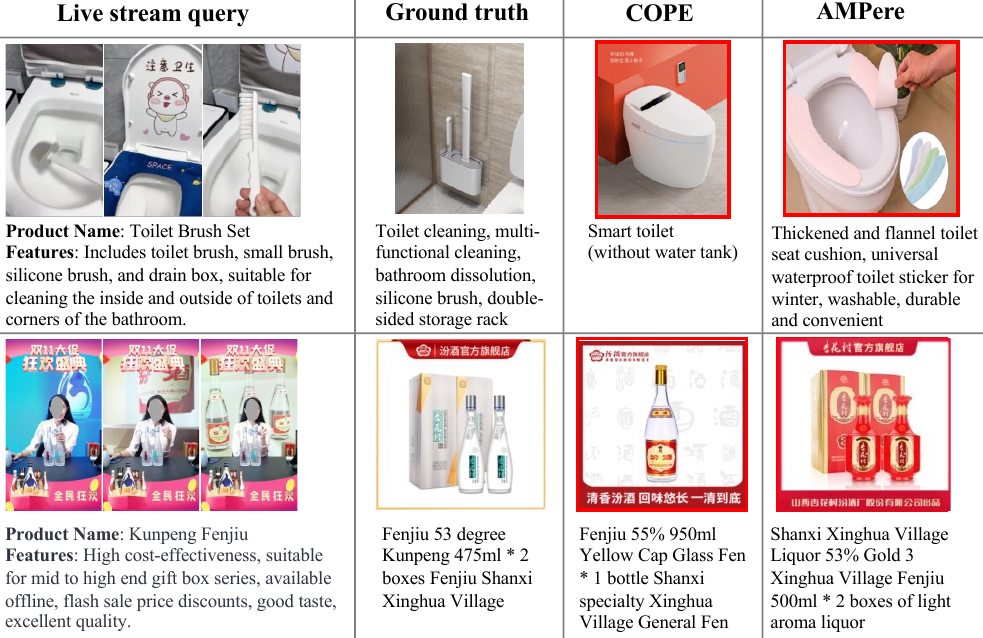}
        \centerline{(a) Failure Type-I: Multi-product presence}
        \label{fig:failed_cases_a}
    \end{minipage}
    \vspace{2mm}
    
    \begin{minipage}[c]{\linewidth}
        \centering
        \includegraphics[width=0.98\textwidth]{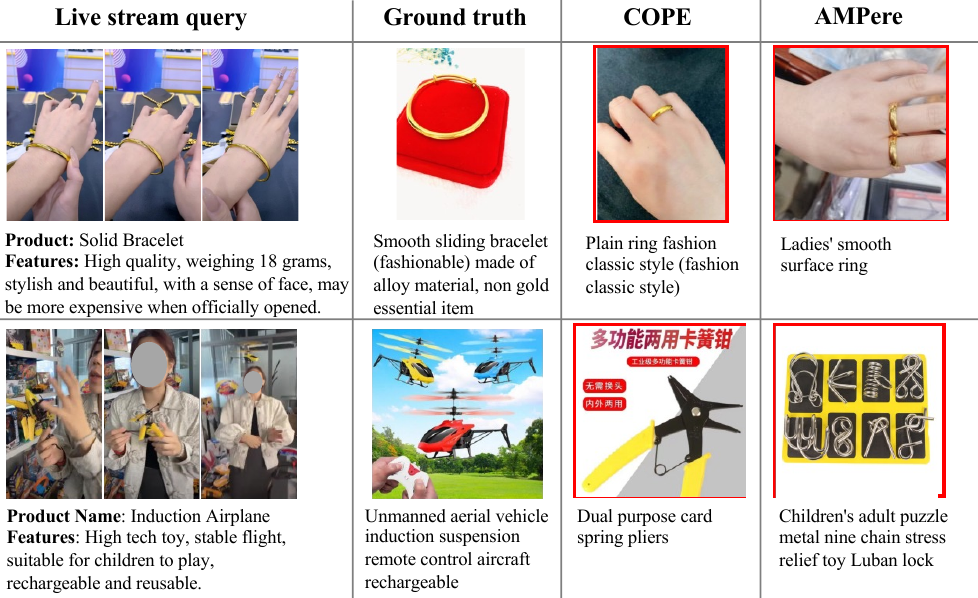}
        \centerline{(b) Failure Type-II: Insufficient multimodal fusion}
        \label{fig:failed_cases_b}
    \end{minipage}
    \vspace{2mm}
    
    \begin{minipage}[c]{\linewidth}
        \centering
        \includegraphics[width=1.0\textwidth]{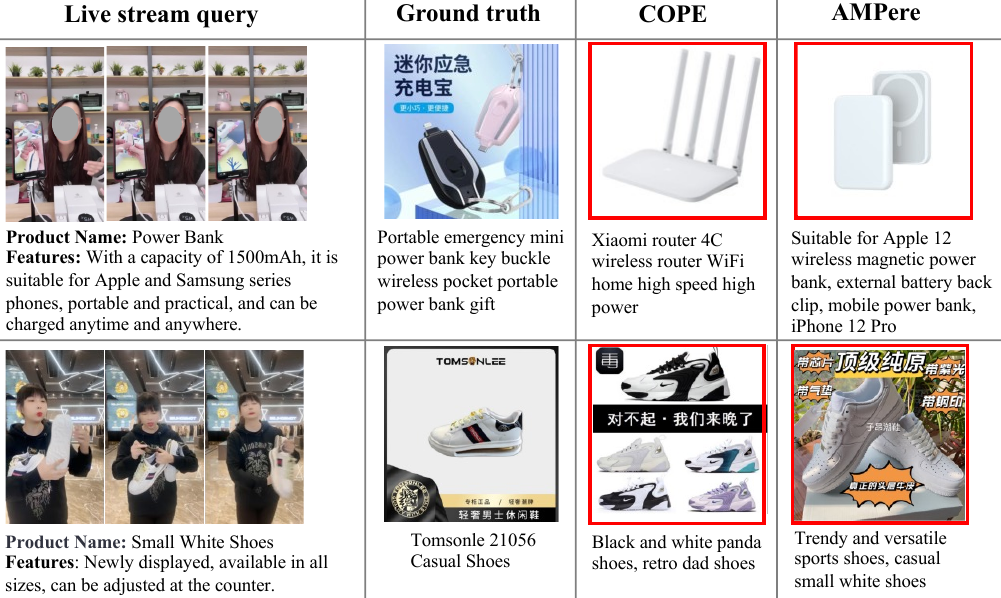}
        \centerline{(c) Failure Type-III: Inaccurate text}
        \label{fig:failed_cases_c}
    \end{minipage}
    \caption{\re{\textbf{Failure cases}. Top-1 item is shown. Task: L2P.}}
    \label{fig:failed_cases}
\end{figure}

\re{\textbf{Failure case analysis}. By manually inspecting the retrieval results of the current worst-performing task, \ie L2P, we observe three common types of errors as follows: \\
$\bullet$ Failure Type-I: \emph{Multi-product presence}. As shown in Fig. \ref{fig:failed_cases_a}, when multiple products co-present in a query video, \eg a toilet brush as the current product and a toilet as a surrounding item, the model may incorrectly retrieve product pages of the surrounding item. To resolve such an ambiguity, product localization is required.  \\
$\bullet$ Failure Type-II: \emph{Insufficient multimodal fusion}.  As illustrated in Fig. \ref{fig:failed_cases_b},  although our LLM based text summarizer has successfully extracted product names, \eg solid bracelet, the result that our model returns as the first hit a photo of a hand with a ring suggests insufficient fusion of the visual and textual inputs. To fully unlock the value of the ASR text, adaptive multimodal fusion requires further research. \\
$\bullet$ Failure Type-III: \emph{Inaccurate text}. This type of failures occurs when the LLM-summarized text, \eg ``power bank'', is not sufficiently accurate to describe the product of ``pocket-portable power bank'' shown in the live stream, see Fig. \ref{fig:failed_cases_c}. How to incorporate the visual context into ASR text summarization for more accurate and fine-grained textual information extraction deserves future exploration.}

\re{\textbf{Efficiency analysis}. The computational cost of training and inference is given in Table \ref{tab:efficiency}. Due to the inclusion of the text branch and the use of LLM, \texttt{AMPere} is relatively slower than COPE. It is worth pointing out that in a practical E-commerce system, instances of specific products are known in advance. Hence, their multimodal embeddings are offline computed and stored for CdPR on the fly. In fact, \texttt{AMPere} has already been deployed in Kuaishou E-commerce for real-time product retrieval and recommendation.}

\begin{table}[htbp]
\centering
\caption{\re{\textbf{Efficiency analysis}. Training 16$\times$V100. Inference on a V100.}}
\label{tab:efficiency}
\setlength{\tabcolsep}{3pt}
\renewcommand{\arraystretch}{1}
\resizebox{\linewidth}{!}{
\begin{tabular}{@{}lrrrrr@{}}
\toprule

\multirow{2}{*}{\textbf{Model}} & 
\multicolumn{2}{c}{\textbf{Training} per epoch} &\multicolumn{3}{c}{\textbf{Inference} per (P,S,L) triplet} \\ 

\cmidrule(r){2-3} \cmidrule(r){4-6} 
& \emph{GPU footprint} & \emph{Time cost} & \emph{GPU footprint} & \emph{LLM} & \emph{Embedding} \\ 
\midrule
COPE & 20.0$\times$16 GB & 22.2 min & 4.6 GB & n.a. & 49.7 ms \\
\texttt{AMPere} & 22.1$\times$16 GB & 31.4 min & 5.1 GB & 4.4 s & 61.7 ms  \\
\bottomrule
\end{tabular}
}
\end{table}

\re{\textbf{Potential ethical issues of using ASR data}. Although E-commerce network anchors are trained to follow professional ethics and codes of conduct, we cannot rule out the possibility that ASR data might occasionally include personally identifiable information, other sensitive, controversial or even offensive content. We recommend the \emph{local} deployment of the LLM-based ASR text summarizer as in our current implementation. However, when a proprietary, cloud-based LLM is preferred, an ethics-oriented pre-filtering on the ASR text is necessary before submitting the text to the remote LLM.}

\section{Conclusions} \label{sec:con}
We propose \texttt{AMPere}, an ASR-enhanced multimodal product representation learning method, for cross-domain product retrieval. Extensive experiments on the large-scale ROPE dataset allow us to draw conclusions as follows. Due to the extremely noisy nature of the ASR text transcribed from short or live-stream videos, directly feeding the raw ASR text to a multimodal representation learning network is suboptimal. LLM is a powerful tool for extracting product-specific information from the verbose text. With the LLM-summarized ASR text, \texttt{AMPere} improves the state-of-the-art visual based solution by a large margin. Finally, parameter sharing across different domains is beneficial for learning a better cross-domain multimodal production representation. Our study shows for the first time that incorporating ASR text can largely benefit product retrieval in the E-commerce domain. As such, we believe this research is likely to inspire further research in similar applications.

\bibliographystyle{IEEEtran}
\bibliography{main}

\begin{IEEEbiography}[{\includegraphics[width=1in,height=1.25in,clip,keepaspectratio]{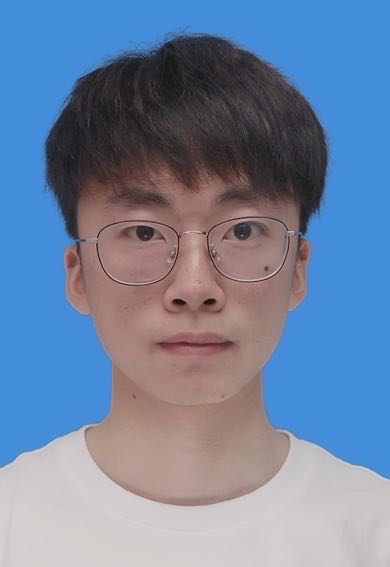}}]{Ruixiang Zhao} received the B.S. degree in Computer Science from Renmin University of China in 2022. He is currently pursuing his  Ph.D. degree at the School of Information, Renmin University of China. His research focuses on multimodal learning and video-text retrieval.
\end{IEEEbiography}
\begin{IEEEbiography}
[{\includegraphics[width=1in,height=1.25in,clip,keepaspectratio]{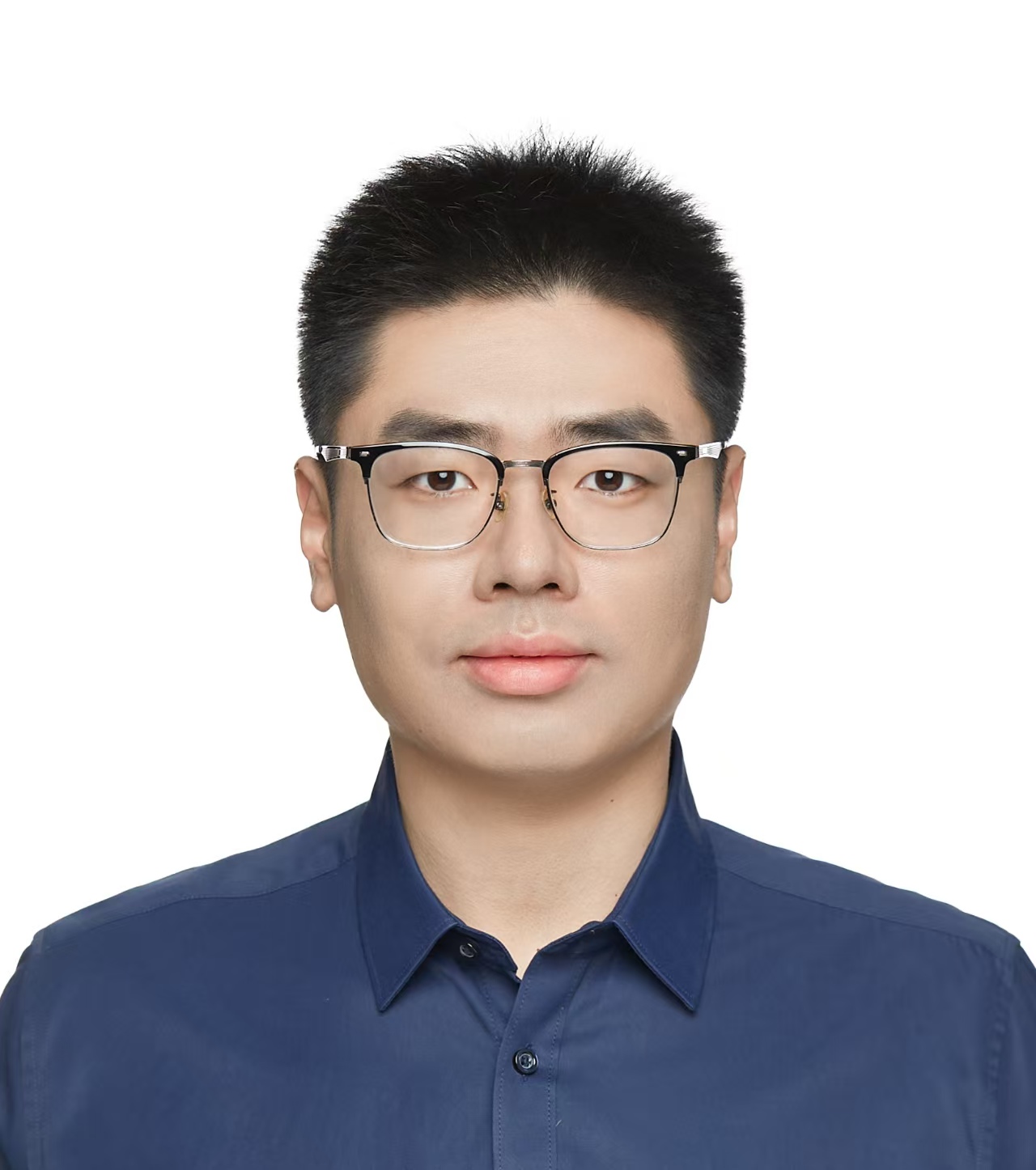}}]{Jian Jia} received the B.E. degree from Shandong University (SDU) in 2015, the M.E. degree from Beijing University of Posts and Telecommunications (BUPT) in 2018, and the Ph.D. degree from University of Chinese Academy of Sciences(UCAS) in 2022. He currently works at Kuaishou, focusing
on research related to computer vision, deep learning, multimodal video understanding.
\end{IEEEbiography}
\begin{IEEEbiography}[{\includegraphics[width=1in,height=1.25in,clip,keepaspectratio]{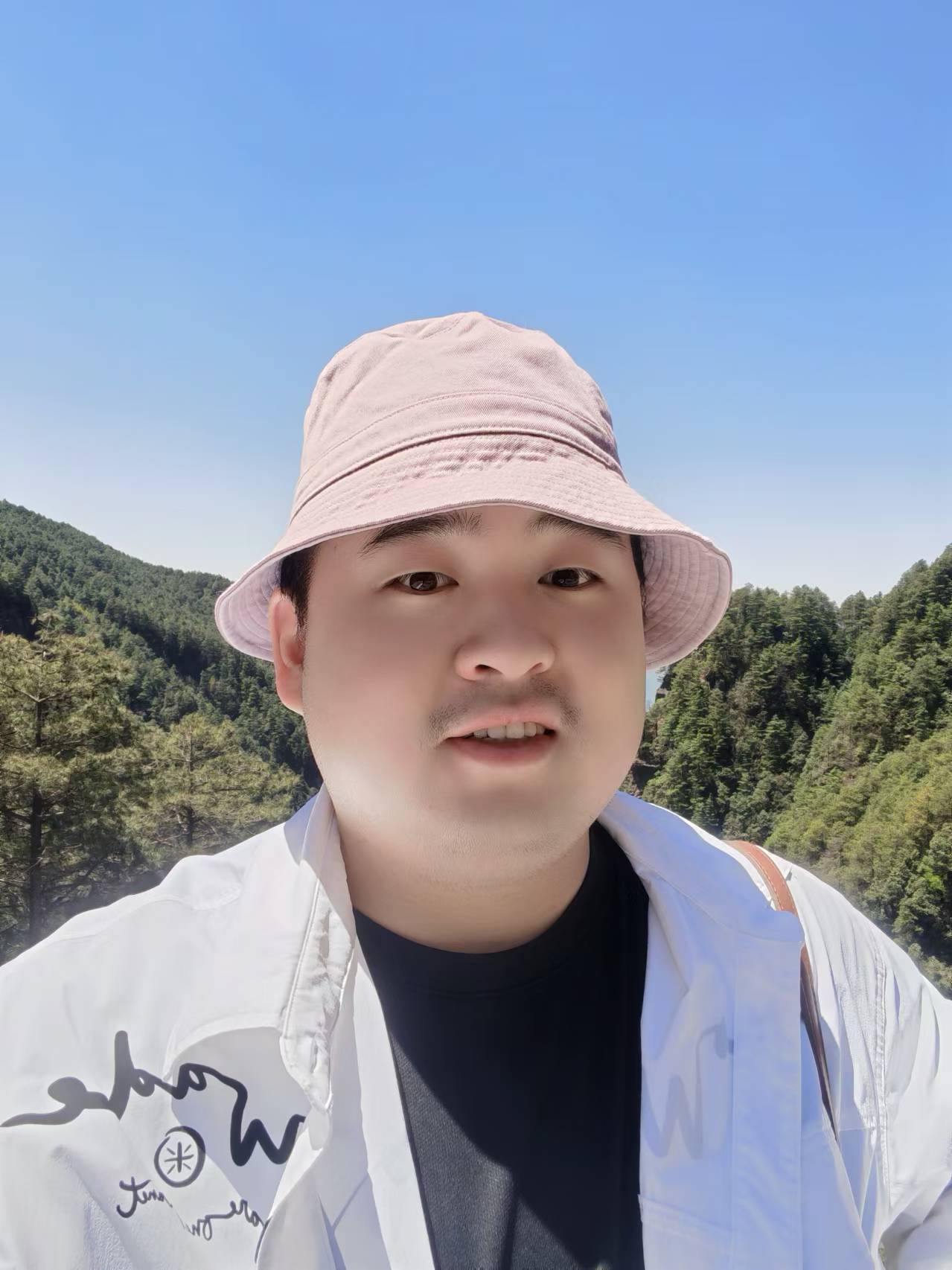}}]{Yan Li} received the B.Eng degree in Automation from Tsinghua University, Beijing, China, in 2014. He received his Ph.D. degree in Pattern Recognition and Intelligent Systems from the Institute of Automation, Chinese Academy of Sciences (CASIA) in 2019. He currently works at TikTok, focusing on research related to multimodal learning and the applications of large language models in recommendation systems.
\end{IEEEbiography}
\begin{IEEEbiography}
[{\includegraphics[width=1in,height=1.25in,clip,keepaspectratio]{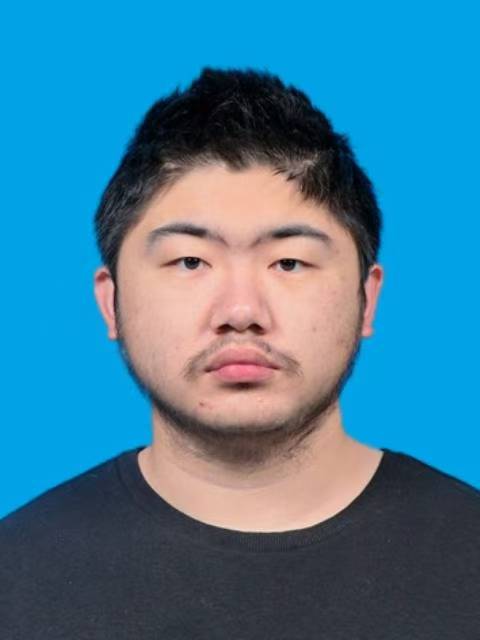}}]{Xuehan Bai}
 received the M.S. degree in information and communication engineering from the Beijing University of Posts and Telecommunications, Beijing, China, in 2022. His research interests include artificial intelligence, large language model, multi-modal large language model.
 \end{IEEEbiography}
\begin{IEEEbiography}
[{\includegraphics[width=1in,height=1.25in,clip,keepaspectratio]{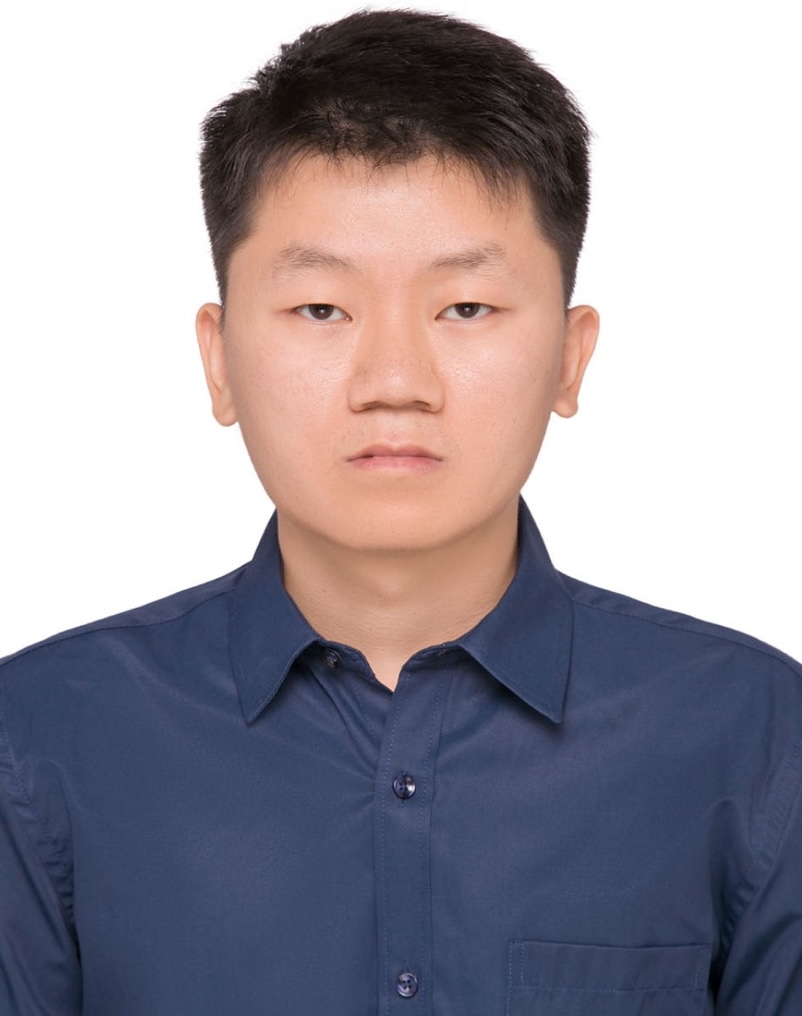}}]{Quan Chen} received the master degree from Beihang University, Beijing, China, in 2015.
He has been a research scientist at Kuaishou Technology since 2022, and before that, he was a research scientist at Alibaba group from 2015 to 2022. His research interests include computer vision, foundation model and online advertising.
\end{IEEEbiography}
\begin{IEEEbiography}
[{\includegraphics[width=1in,height=1.25in,clip,keepaspectratio]{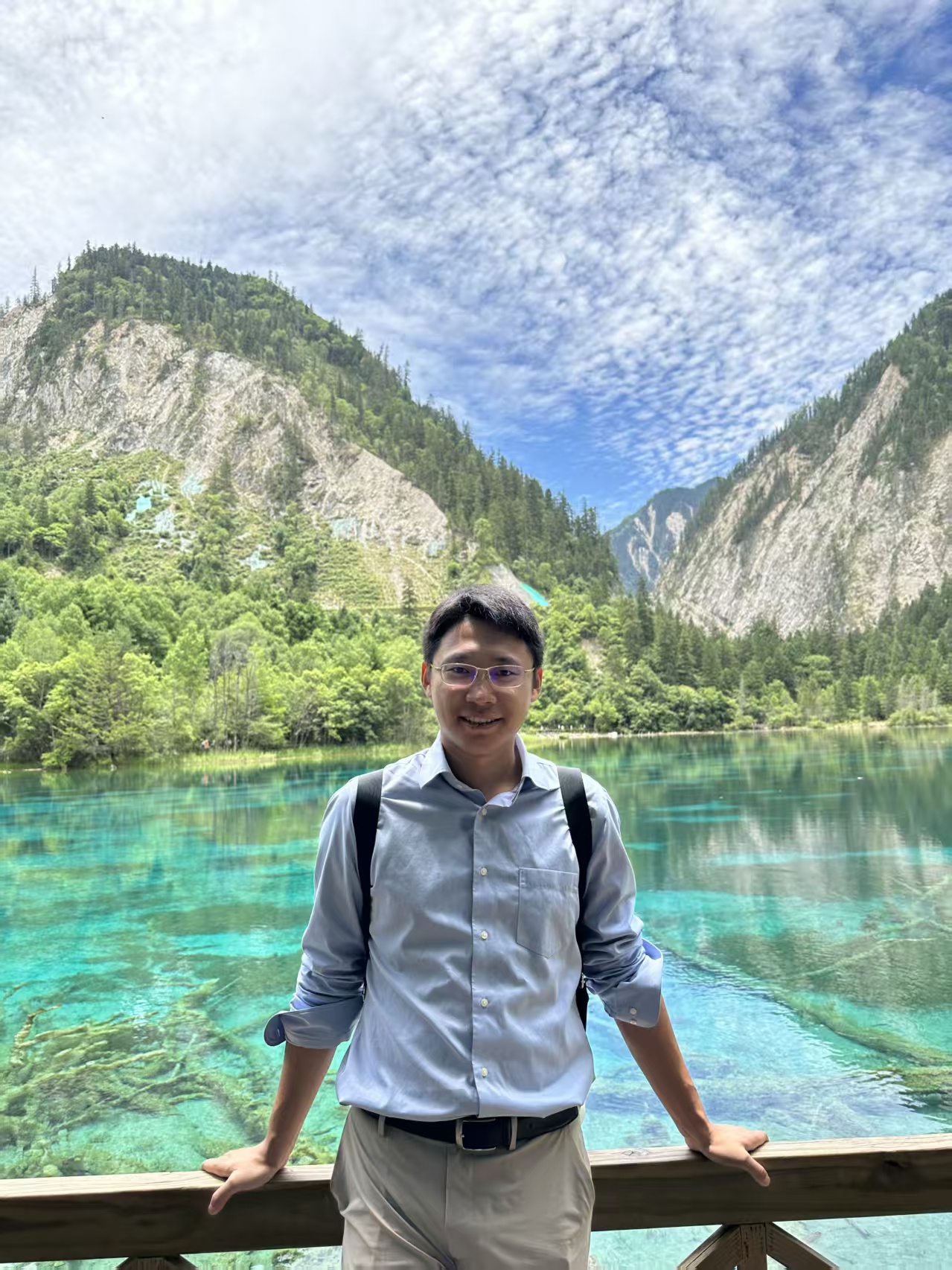}}]{Han Li} received the B.S. and M.E. degrees from Tsinghua University, Beijing, China, in 2010 and 2013, respectively. He is currently VP of Kuaishou Inc., leading the short-video recommendation algorithm team. Before joining Kuaishou, he was a director of Alibaba  advertising team. He has published dozens of papers in several top conferences, such as, NeuralIPS, ICML, KDD, AAAI, WWW, ICCV. His research interests include machine learning, online advertising and recommendation system.
\end{IEEEbiography}
\begin{IEEEbiography}
[{\includegraphics[width=1in,height=1.25in,clip,keepaspectratio]{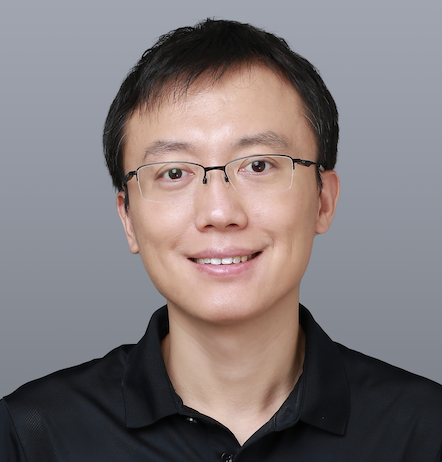}}]{Peng Jiang} is VP of Kuaishou. He is leading the Advertising Algorithm team to develop machine learning algorithms that power Kuaishou's ads engines, including both domestic and international product (Kuaishou and Kwai). Before joining Kuaishou, he was a director in Alibaba, leading the recommendation team. He has over 70 publications appeared in several top conferences such as KDD, SIGIR, WWW, NIPS, ICLR, and journals including TOIS. He received the Best Paper Award of CIKM 2022, the Best Paper Award Honourable Mention of SIGIR 2023 and the Best Paper Runner Up of RecSys 2019. He received the Ph.D. degree from Beijing Institute of Technology in 2012.
\end{IEEEbiography}
\begin{IEEEbiography}
[{\includegraphics[width=1in,height=1.25in,clip,keepaspectratio]{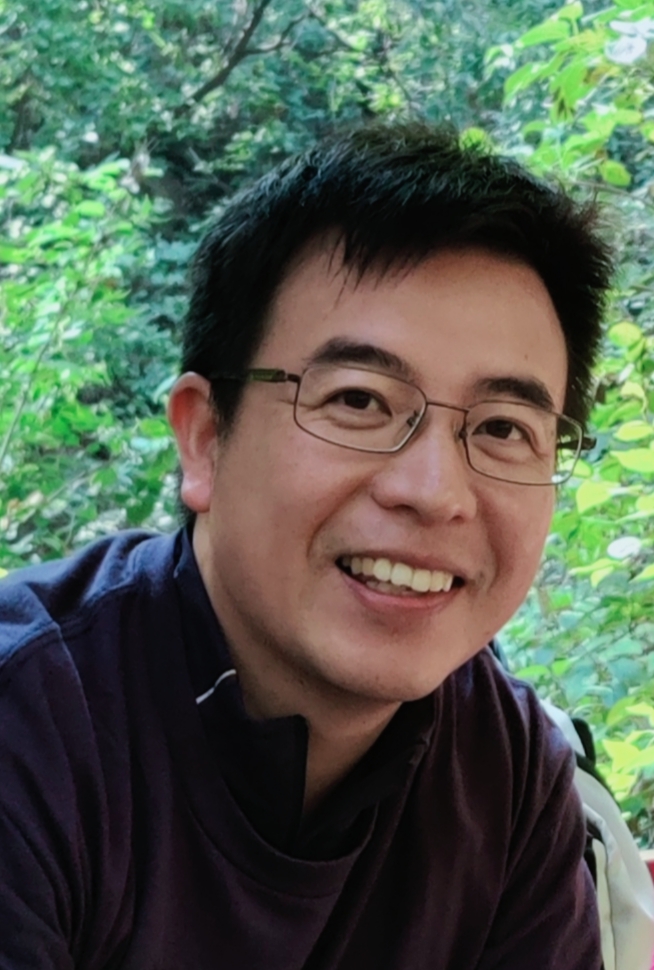}}]{Xirong Li} (Member, IEEE) received the B.S. and M.E. degrees from Tsinghua University, Beijing, China, in 2005 and 2007, respectively, and the Ph.D. degree from the University of Amsterdam, Amsterdam, The Netherlands, in 2012, all in computer science. He is currently a Full Professor with the School of Information, Renmin University of China, Beijing. His research focuses on multimodal intelligence. He was recipient of CCF Science and Technology Award 2024, the ACMMM 2016 Grand Challenge Award, the ACM SIGMM Best Ph.D. Thesis Award 2013, the IEEE Transactions on Multimedia Prize Paper Award 2012, and the Best Paper Award of ACM CIVR 2010. He served as the Program Co-Chair for Multimedia Modeling 2021 and is serving as an Associate Editor for ACM TOMM and the Multimedia Systems journal.
\end{IEEEbiography}

\vspace{11pt}

\vfill
\clearpage\end{CJK*}

\end{document}